\documentclass{aa}  
\usepackage{graphicx}
\usepackage{txfonts}
\usepackage{natbib}
\usepackage{longtable}
\usepackage{hyperref}
\bibpunct{(}{)}{;}{a}{}{,}   
\def\kms{km\,s$^{-1}$}
\begin{document}
   \title{A comprehensive chemical abundance study of the outer halo globular cluster M~75
}

%
   
\author{N.~Kacharov\inst{1}\thanks{Member of the International Max Planck Research School for Astronomy and Cosmic Physics at the University of Heidelberg, IMPRS-HD, Germany.}
	\and A.~Koch\inst{1}
        \and A.~McWilliam\inst{2}
	}

   \offprints{N. Kacharov, n.kacharov@lsw.uni-heidelberg.de}

   \institute{Landessternwarte, Zentrum f\"{u}r Astronomie der Universit\"{a}t Heidelberg, K\"{o}nigstuhl 12, D-69117 Heidelberg, Germany
         \and
              Carnegie Observatories, 813 Santa Barbara St., Pasadena, CA 91101, USA    
             }

   \date{Received 03.03.2013 / Accepted 11.04.2013}

 
  \abstract
   {M~75 is a relatively young Globular Cluster (GC) found at $15$~kpc from the Galactic centre at the transition region between the inner and outer Milky Way halos.} 
   {Our aims are to perform a comprehensive abundance study of a variety of chemical elements in this GC such as to investigate its chemical enrichment history in terms of early star formation, and to search for any multiple populations.}
   {We have obtained high resolution spectroscopy with the MIKE instrument at the Magellan telescope for $16$ red giant stars. Their membership within the GC is confirmed from radial velocity measurements. Our chemical abundance analysis is performed via equivalent width measurements and spectral synthesis, assuming local thermodynamic equilibrium (LTE).}
   {We present the first comprehensive abundance study of M~75 to date. The cluster is metal-rich ([Fe/H]~$=-1.16\pm0.02$~dex, [$\alpha$/Fe]~$=+0.30\pm0.02$~dex), and shows a marginal spread in [Fe/H] of $0.07$~dex, typical of most GCs of similar luminosity. A moderately extended O-Na anticorrelation is clearly visible, likely showing three generations of stars, formed on a short timescale. Additionally the two most Na-rich stars are also Ba-enhanced by $0.4$ and $0.6$~dex, respectively, indicative of pollution by lower mass ($M\sim4-5~M_{\odot}$) Asymptotic Giant Branch (AGB) stars. The overall n-capture element pattern is compatible with predominant r-process enrichment, which is rarely the case in GCs of such a high metallicity.} 
  {}


   \maketitle
%

\section{Introduction}


With ages of $10 - 14$~Gyr, globular clusters (GCs) are the oldest stellar systems in the Galaxy and therefore reflect its earliest evolutionary stages. Although the Galactic halo GC system appears homogeneous \citep[e.g.][]{cohen+melendez2005,koch+2009} and well compatible with the stellar halo in many regards \citep[e.g.][]{geisler+2007}, numerous properties show broad differences between individual clusters and are at odds with halo field stars.
These characteristics comprise large spreads and anti-correlations of the light- elements involved in p-capture reactions \citep[C, N, O, F, Na, Mg, and Al,][]{osborn71,cottrell+dacosta81,kraft94,gratton+2001,gratton+2004,carretta+2009c,carretta+2009b}.
Nowadays, these variations amongst the light elements are considered as an evidence for the existence of at least two generations of stars, present in all GCs studied to date \citep[e.g.][and references therein]{gratton+2012},  as also often prominently seen in their colour-magnitude diagrams \citep[e.g.][]{piotto+2012}. The multiple populations in GCs are tightly linked to the ``second-parameter effect'', which needs to explain discordant horizontal branch (HB) morphologies at any given metallicity. Suggested solutions to this problem include a broad age range in the GCs \citep{searle+zinn78} or variations in their helium content \citep[e.g.][]{dantona+2002}, whilst mass loss, $\alpha$-abundances, rotation, deep mixing, binary interactions, core concentration, or planetary systems cannot be ruled out as possible second parameters \citep[see][for a detailed review]{catelan2009}.

Currently, there are several theories trying to explain the formation of at least two stellar populations in GCs. The best candidates that pollute the interstellar medium (ISM) with p-capture elements whilst producing only little or no $\alpha$- and Fe-peak-elements are massive ($\sim5 - 8 M_{\odot}$) AGB stars \citep{dercole+2008,dercole+2010} or fast rotating massive ($M>10 M_{\odot}$) stars \citep[FRMS][]{decressin+2007}. Both mechanisms work on very different timescales: The winds of FRMS enrich the ISM with p-capture products in $\sim6\times10^6$~yrs, slightly before the explosions of the bulk of SNe II take place. 
On the other hand, the long-lived AGB stars take a few $10^8$~yrs before they enrich the ISM with these elements \citep{gratton+2012}. A recent study by \citet{valcarce+catelan2011} attempts to combine both mechanisms. A problem in all theories is the small fraction of primordial first generation (FG) stars ($\sim30\%$) with respect to the second generation (SG) stars ($\sim70\%$) in present-day GCs, since the observed mass of the FG is not sufficient to form the numerous SG by a factor of $\sim10$ \citep{dercole+2008}.
For both mechanisms to work, one has to either invoke a top-heavy IMF for the FG or to assume that the GCs were much more massive and they lost a large fraction of their initial mass. \citet{deMink+2009} made an attempt to solve this problem by suggesting massive binaries as the main polluters.
In fact, the similarity in chemistry and ages of the Milky Way halo field stars with the FG stars in GCs and the discovery of a few halo stars with modified CN and CH abundances, suggest that the bulk of the halo stars were indeed formed in GCs \citep{gratton+2012,martell+grebel2010} and that the GCs originated in the centres of much larger and later disrupted stellar systems \citep[e.g.][]{carretta+2010,boker2008,kravtsov+gnedin2005}.

Coupled with the lack of a metallicity gradient in the outer halo GC system, the outer halo clusters' second parameter problem had prompted the first suggestion by \citet{searle+zinn78} that those GCs could have been donated by accreted dwarf galaxy-like systems. The most distant Milky Way GCs present a number of properties, which suggest a different origin than the inner halo GCs \citep{rodgers+paltoglou84,zinn93,zinn96,marin-franch+2009}. These comprise younger ages, different kinematics, and possibly different chemical composition. Therefore, studies of GCs at larger Galactic distances are crucial for the understanding of how the Galactic halo formed.

In this paper we present the first ever chemical element abundances derived from high-resolution spectra for the GC M~75. This cluster is located at a galactocentric distance of $15$~kpc, which tenants the transition region between the inner and outer Milky Way halo \citep{zinn93,carollo+07}. Its younger age \citep[$\sim10$~Gyr;][]{catelan+2002} and high metallicity ([Fe/H]$=-1.16$~dex, this work) are compatible with the properties of the outer halo GC system and suggest a possible extragalactic origin. On the other hand, M~75 is amongst the most concentrated GCs ($c=\log(r_t/r_c)=1.80$), which could be contrasted to the extended and loose clusters in the outer halo \citep{koch+cote2010,koch+2009}. This unique GC also has a trimodal horizontal branch (HB), the origin of which is not explicable under canonical stellar evolutionary models \citep{catelan+2002}.
Apart from the well separated red HB (RHB) and blue HB (BHB), its CMD shows a distinct third extension of a very blue, faint tail. Moreover, it has an 
anomalously low ratio of red giant branch (RGB) to HB stars, indicating higher He-content of the cluster. Thus, it is very important to assess possible multiple populations, which could be related to the peculiar HB morphology and to look for peculiarities in its chemical composition, which might reveal clues for its origin and early evolution.

\section{Observations and data reduction}

\begin{figure}
\centering
\resizebox{\hsize}{!}{
\includegraphics[angle=0]{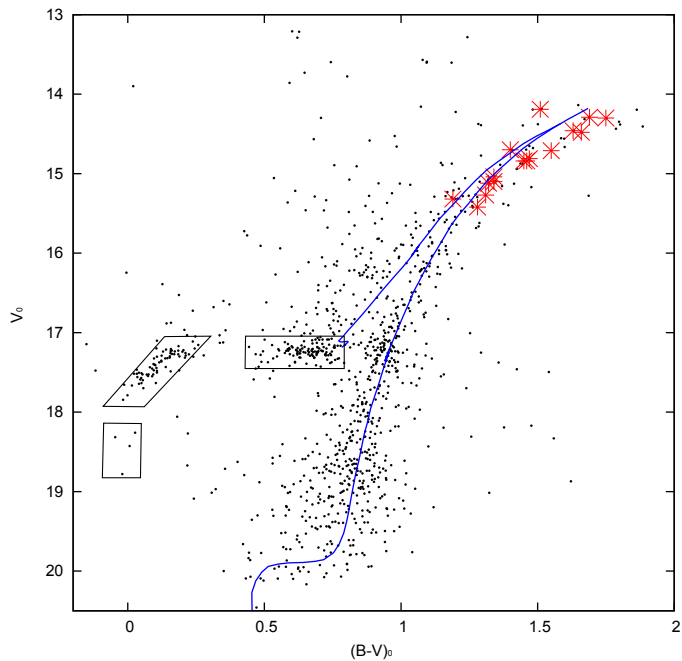}
}
\caption[]{CMD from \citet{kravtsov+2007}. The stars from our sample are indicated by larger red symbols. An isochrone for [Fe/H]~$\sim-1.2$~dex and age of 10 Gyr from the Padova library \citep{marigo+08,girardi+10} is overplotted for comparison. The three HBs are indicated by black rectangles.
}
\label{fig:CMD}
\end{figure}

Our spectroscopic observations of 16 giant stars in M~75 were taken using the Magellan Inamori Kyocera Echelle (MIKE) spectrograph at the 6.5-m Magellan2/Clay Telescope at Las Campanas Observatory, Chile. The instrument consists of two arms sensible in the red and blue parts of the visible spectrum, which cover an entire wavelength range of $3340$~\AA ~to $9150$~\AA. Our data were collected over one night in April and four nights in July 2011. By using a slit width of $0.7\arcsec$ and $2\times1$ binning of the CCD in spatial and spectral direction, we obtained a spectral resolution of approximately $30000$. The typical seeing during the runs was $\sim1\arcsec$ on average. We reached a relatively high S/N of $\sim70$ per pixel around $6500$~\AA~ on the red CCD and $\sim40$ per pixel around $4500$~\AA~ on the blue CCD. The observing log is presented in Table \ref{tab:Obs_log}.
The targets were selected from the catalogue of \citet{kravtsov+2007}, choosing stars with a high membership probability, i.e. those within the tidal radius of the cluster, yet avoiding the crowded central regions. This was aided by visual inspection of archival FORS preimaging (Program ID 69.B-0305, P.I. E. Tolstoy).
A colour-magnitude diagram (CMD) of M~75, highlighting our spectroscopic sample, is presented in Figure \ref{fig:CMD}. We also overplotted an isochrone of age $10$~Gyr and metallicity $Z = 0.003$ from the Padova library \citep{girardi+10,marigo+08}, which best represents the photometric data. This adopts an extinction value $A_V=0.49$~mag ($E(B-V)=0.147$~mag) from \citet{schlegel+98}, and a distance modulus of $(m-M)_0 = 16.4$~mag (linear distance $19$~kpc; \citet{catelan+2002}). Our sample consists of $13$ RGB stars and $3$ possible AGB stars.

The data were processed with the MIKE pipeline reduction package \citep{kelson+2000,kelson2003}, which comprises flat field division, order tracing from quartz lamp flats, and wavelength calibration using built-in Th-Ar lamp exposures that were taken immediately following each science exposure. Continuum normalisation was performed by dividing the extracted spectra by a high-order polynomial fit to a spectrum of an essentially line-free hot rotating star, taken as part of our observing runs.

\begin{table}
\begin{center}
\caption{Observing Log.}\label{tab:Obs_log}
{\small
 \begin{tabular}{cccc}
\hline
 Star ID\tablefootmark{1} & V$_0$ & Date   &  Exp. time \\
         & [mag]   &        &   [s]      \\
\hline
239   & 15.04	&  Jul. 25 2011  &     $3\times1600$  \\
251   & 14.48	&  Jul. 23 2011  &     $3\times900$  \\
442   & 14.81	&  Jul. 26 2011  &     $1200+518$  \\
461   & 14.29	&  Jul. 26 2011  &     $2\times800$ \\
483   & 14.84	&  Jul. 24 2011  &     $3\times1200$  \\
486   & 15.27	&  Jul. 25 2011  &     $3\times2000$ \\
503   & 14.30	&  Jul. 23 2011  &     $1\times2700$  \\
512   & 15.12	&  Jul. 25 2011  &     $3\times1600$  \\
583   & 14.19	&  Apr. 04 2011  &     $2400$  \\
612   & 15.32	&  Jul. 26 2011  &     $3\times2000$ \\
655   & 14.84	&  Jul. 24 2011  &     $3\times1200$  \\
876   & 15.42	&  Jul. 26 2011  &     $2000+1600+3\times900$  \\ 
901   & 14.71	&  Jul. 24 2011  &     $3\times1200$ \\
1251  & 15.10	&  Jul. 24 2011  &     $1600 + 1541$  \\
1312  & 14.46	&  Jul. 23 2011  &     $3\times900$  \\
1459  & 14.70	&  Jul. 23 2011  &     $3\times1200$  \\
\hline
 \end{tabular}
\par}
\tablefoot{
\tablefoottext{1}{Based on the catalogue of \citet{kravtsov+2007}.}
}
\end{center}
\end{table}

We computed the radial velocities of our targets by crosscorrelating the spectra with a synthetic RGB spectrum with stellar parameters similar to the target stars using the $fxcor$ tool in IRAF. From this, we found a mean heliocentric velocity of $-186.2\pm1.9$~\kms (standard deviation $8.1$~\kms), confirming the cluster membership of all stars. This is in an excellent agreement with the mean systematic radial velocity of M~75 of $-189.3$~\kms, $\sigma = 10.3$~\kms \citep[2010 version]{harris96}. Whilst this dispersion may seem large for a GC, it has to be kept in mind that M~75 is a very massive and concentrated system, so that this value is fully compatible with the large velocity dispersions found in other comparably luminous systems \citep{pryor+meylan93}.

\section{Abundance analysis}

\subsection{Line list}

We derived chemical element abundances through an equivalent width (EW) analysis, complemented by spectral synthesis using the stellar abundance code MOOG \citep{sneden73}. We used an absolute abundance analysis method, which closely follows the procedures described in \citet{koch+2009,koch+cote2010}.
The line list was assembled from various sources \citep[][and references therein]{koch+cote2010}, and complemented with atomic data from the Kurucz data base\footnote{\url{http://www.cfa.harvard.edu/amp/ampdata/kurucz23/sekur.html}}. Additional transitions for some heavier elements were adopted from \citet{sadakane+2004} and \citet{yong+2008}.
The EWs of the lines were measured by fitting Gaussian profiles to the absorption features using the $splot$ task in IRAF.
We restricted our measurements to lines having reduced EWs ($\log($EW$/\lambda)$) less than $-4.5$ to avoid the saturated parts of the curves of growth. A major source of random error in the EWs estimates, especially in the bluer regions (below $5500$~\AA), comes from difficulties in placing the continuum due to the strong blending owing to the relatively high metallicity of M75. We used the deblending option in IRAF's task $splot$ to account for blended lines where necessary.

For the elements Rb, Zr, Gd, Dy, Er, Hf, and Th, we used spectral synthesis instead. Accurate EW measurements were not possible because of strong blending, too weak lines, or too low S/N ratio.

We applied corrections for Hyperfine Structure (HFS) splitting for the odd-Z elements V, Mn, Co, Cu, Rb, La, and Eu using the $blends$ driver of MOOG and atomic data for the splitting from \citet[][in prep.]{mcwilliam+95,mcwilliam+2012}. HFS corrections for Sc and Ba were small compared to the $1\sigma$ measurement uncertainty and we ignored them. HFS corrections for the lighter odd-Z elements are generally negligible.

Finally, the derived abundances were placed on the solar scale of \citet{Asplund+2009}.
The full linelist and the measured EWs are available in the online version of A\&A. We present the first rows and columns of this table to guide the eye (Table \ref{tab:linelist}).

\begin{table*}
\begin{center}
\caption{Line list and equivalent widths. The full table is available in the electronic version of the journal.}\label{tab:linelist}
{\footnotesize
 \begin{tabular}{cccccccccccccccc}
\hline
Element & $\lambda$ & $\chi$ & $\log$~gf & \multicolumn{12}{c}{EWs [m\AA] for each star} \\
        & [\AA]     & [eV]  &           & \#239 & \#251 & \#442 & \#461 & \#483 & \#486 & \#503 & \#512 & \#583 & \#612 & \#655 & ... \\ 
\hline
$\rm{[O   I]}$ & 6300.31 &  0.00 & -9.819   &	58 &	63 & 30  &   100 &    31   &	21  &	 59  &  51  &	53  &	...  &   50  &    \\
$\rm{[O   I]}$ & 6363.79 &  0.02 & -10.303  &	23 &	27 & 18  &   47  &    13   &	 6  &	 32  &  31  &	42  &	...  &	 22  &    \\
Na  I & 5682.63 &  2.10 & -0.700   &	86 &   142 & 141 &   111 &    143  &	109 &	 131 &	 82 &	105 &	  98 &	 118 &   \\
Na  I & 6154.23 &  2.10 & -1.560   &	19 &	58 & 64  &   40  &    73   &	43  &	 61  &  28  &	36  &	 33  &	 39  &    \\
Na  I & 6160.75 &  2.10 & -1.260   &	31 &	77 & 82  &   64  &   102   &	66  &	 79  &  26  &	49  &	 44  &	 63  &    \\
Mg  I & 5528.42 &  4.35 & -0.357   &   199 &   247 & 214 &   233 &    221  &	206 &	 243 &  199 &	227 &	 206 &	 223 &   \\
Mg  I & 5711.09 &  4.33 & -1.728   &   113 &   147 & 129 &   126 &    127  &	122 &	 136 &  111 &	124 &	 104 &	 124 &   \\
Al  I & 6696.03 &  3.14 & -1.347   &	23 &	69 & 87  &   50  &    76   &	54  &	 47  &  24  &	46  &	 40  &	 26  &    \\
Al  I & 6698.67 &  3.14 & -1.647   &   ... &    32 & 40  &   20  &    39   &    26  &    24  &  ... &   20  &    22  &   23  &   \\
Al  I & 7835.31 &  4.02 & -0.649   &    13 &    39 & 46  &   30  &    52   &    38  &    35  &  17  &   23  &    30  &   22  &   \\
Al  I & 7836.13 &  4.02 & -0.494   &    17 &    49 & 56  &   33  &    60   &    47  &    36  &  20  &   26  &    35  &   26  &   \\
Si  I & 5684.48 &  4.95 & -1.650   &	46 &	55 & 56  &   40  &    70   &	56  &	 47  &  45  &	50  &	 49  &	 47  &    \\
Si  I & 5948.55 &  5.08 & -1.230   &	69 &	67 & 79  &   69  &    81   &	72  &	 60  &  66  &	57  &	 70  &	 68  &    \\
Si  I & 6155.13 &  5.61 & -0.750   &	49 &	48 & 50  &   55  &    49   &	48  &	 50  &  55  &	45  &	 53  &	 51  &    \\
Ca  I & 5261.71 &  2.52 & -0.580   &   126 &   173 & 152 &   175 &    144  &	121 &	 158 &  120 &	153 &	 114 &	 139 &   \\
Ca  I & 5590.13 &  2.52 & -0.570   &   124 &   166 & 139 &   166 &    145  &	120 &	 140 &  120 &	134 &	 114 &	 123 &   \\
Ca  I & 5601.29 &  2.53 & -0.520   &   131 &   183 & 163 &   188 &    145  &	127 &	 175 &  133 &	171 &	 120 &	 148 &   \\
Ca  I & 5857.46 &  2.93 &  0.230   &   147 &   182 & 166 &   150 &    159  &	142 &	 176 &  146 &	163 &	 128 &	 154 &   \\
Ca  I & 6166.44 &  2.52 & -1.140   &	96 &   129 & 113 &   141 &    120  &	 99 &	 133 &	 97 &	120 &	  77 &	 106 &	 \\
Ca  I & 6169.04 &  2.52 & -0.800   &   120 &   152 & 145 &   144 &    133  &	126 &	 150 &  114 &	139 &	 102 &	 132 &   \\
Ca  I & 6169.56 &  2.52 & -0.480   &   132 &   169 & 146 &   179 &    150  &	139 &	 173 &  134 &	158 &	 128 &	 146 &   \\
Ca  I & 6455.60 &  2.52 & -1.290   &	89 &   110 & 100 &   112 &    101  &	 92 &	 117 &	 80 &	110 &	  70 &	  92 &	 \\
Ca  I & 6471.67 &  2.52 & -0.875   &   122 &   162 & 149 &   157 &    151  &	126 &	 162 &  124 &	154 &	 118 &	 138 &   \\
Ca  I & 6499.65 &  2.52 & -0.820   &   123 &   155 & 138 &   160 &    135  &	123 &	 149 &  116 &	145 &	 100 &	 130 &   \\
Ca  I & 6717.69 &  2.71 & -0.610   &   146 &   193 & 170 &   188 &    170  &	152 &	 188 &  142 &	180 &	 137 &	 166 &   \\
...   &         &       &          &       &       &     &       &         &        &        &      &       &        &       &      \\
\hline 
 \end{tabular}
\par}
\end{center}
\end{table*}

\subsection{Stellar atmospheres}

We interpolated the new grid of Kurucz\footnote{\url{http://www.cfa.harvard.edu/grids.html}} plane-parallel, one-dimensional models without convective overshoot. These include the $\alpha$-enhanced opacity distribution functions \citep[AODFNEW;][]{castelli+kurucz2003}\footnote{\url{http://wwwuser.oat.ts.astro.it/castelli}}.

\begin{figure}
\centering
\resizebox{\hsize}{!}{
\includegraphics[angle=0]{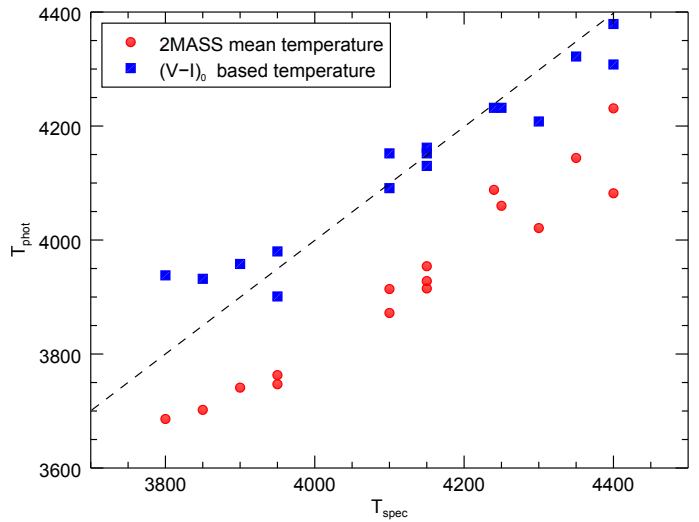}
}
\caption[]{A comparison between the spectroscopic and photometric temperature estimates. The average temperature calculated from $(V-J)_0$, $(V-H)_0$, and $(V-K)_0$ colours is indicated by red circles and the $(V-I)_0$ calibration is indicated by blue squares. The dashed line is unity.  
}
\label{fig:temp}
\end{figure}

As an initial guess, we calculated effective temperatures of our targets based on the $(V-I)_0$ colours from the photometric catalogue of \citet{kravtsov+2007}. This was complemented with photometry from 2MASS \citep{cutri+03} to obtain temperature-estimates based on the $(V-J)_0$, $(V-H)_0$, and $(V-K)_0$ colour indices. We used the temperature-colour calibrations of \citet{ramirez+melendez2005}.
Additionally, we obtained spectroscopic temperatures by measuring the EWs of a large number, typically about $60$, Fe I lines and establishing excitation equilibrium. This is achieved by changing the temperature until there is no correlation between the derived abundances from different Fe I lines and their excitation potential. As a result, the mean temperature from all three 2MASS based indicators is lower than the temperature based on the $(V-I)_0$ colour alone and the spectroscopic estimates by $200$~K on average (Figure \ref{fig:temp}). A similar trend was also noted by \citet{fabbian+2009}. One possible explanation is the larger pixel size of the 2MASS detectors, which can lead to an undersampling in the crowded GC field compared to better sampled optical images.
This way, additional flux contributions per pixel would yield overestimated infrared magnitudes and thus lower effective temperatures. We use the spectroscopic temperatures in the following analysis. The mean difference between the temperatures from the $(V-I)_0$ colours and the spectroscopic ones is only $2$~K with a $1\sigma$-scatter of $60$~K, which we adopted as the temperature error for our targets.
  
\begin{figure}
\centering
\resizebox{\hsize}{!}{
\includegraphics[angle=0]{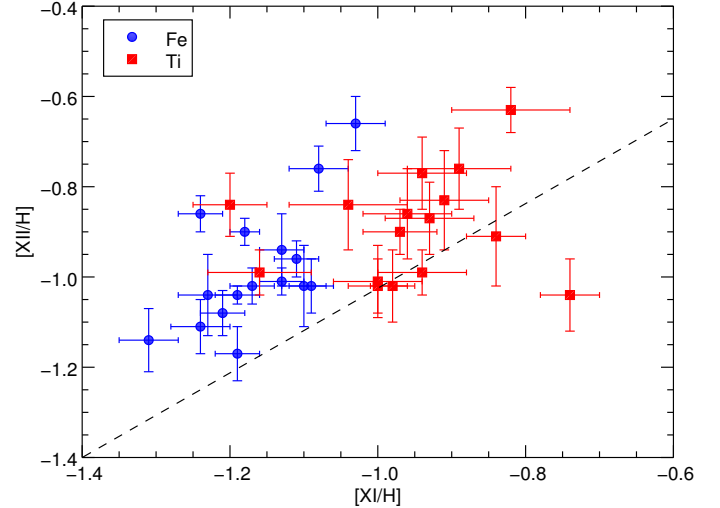}
}
\caption[]{A comparison between the abundance results from neutral and ionised species. Fe abundances are shown with blue circles and Ti abundances with red squares. The dashed line is unity.  
}
\label{fig:ion}
\end{figure}

We derived physical gravities from the canonical Equation 1, using the dereddened $V_0$ magnitudes with a bolometric correction interpolated from the Kurucz grid, the spectroscopic temperatures, and adopting the known distance to the cluster ($19$~kpc). We adopted a mass $\mu = 0.78$~M$_{\odot}$ for all stars, which is consistent with the masses from the reference isochrones. Adopting lower masses for the possible AGB stars in our sample would lead to a small change in gravity, which would have a negligible effect on the derived abundances (See Section 3.3).
\begin{equation}
 \log g = \log(\mu/\mu_\odot) + 4\log(T/T_\odot) - 0.4(M_\odot - M) + \log g_\odot
\end{equation}
In the above equation, $M=M_V-BC$ denotes the absolute bolometric magnitude of the stars. We did not adjust the gravities to enforce ionisation equilibrium. As a result, the abundances from the Fe II lines are higher by $0.18$ dex ($\sigma = 0.10$~dex) on average, compared to the neutral species (Figure \ref{fig:ion}). We note that the differences are larger for cooler stars, which might be due to the use of plane-parallel models instead of spherical or 3D ones \citep{bergemann+2012}, departure from LTE \citep{heiter+eriksson2006,bergemann+2012,ruchti+2013}, or unknown blends.
Additionally, the use of iron lines with a broad range of excitation potentials (from $1$ to $5$~eV) could also cause some discrepancy, as noted by \citet{worley+2010} and \citet{worley+cottrell2010}.
The discrepancy in Fe I vs. Fe II is too large to be explained by systematic errors. Shifting the gravities by $0.5$~dex would restore ionisation equilibrium, but this implies more than a $50\%$ error in the distance to the cluster, which is rather unlikely. A change in the interstellar extinction towards M~75 by $\Delta E(B-V)=0.35$~mag or an increase of the temperature scale by $100$~K ($200$~K for the coolest stars) will also restore ionisation equilibrium. We deemed both possibilities unlikely, given the large, required changes compared to the small uncertainties in the parameters and our overall excellent excitation equilibrium.
Likewise, the Ti abundances from the ionised species are larger than the ones based on Ti I lines by $0.07$~dex ($\sigma = 0.15$~dex).

Microturbulent velocities ($\xi$) were determined by removing any trend in the plot of abundances versus EW of the Fe I lines. The derived values for M~75 are in the order of $2.1$~\kms. A typical error of this method is $\sim0.1$~\kms: variations within this range still allow reasonably flat slopes in the EW plot. 

Since we did not have a prior knowledge of the metallicities of the individual stars, we started with atmosphere models for [M/H]~$=-1.2$~dex, as representative for the cluster mean \citep{catelan+2002}, and updated it iteratively based on the Fe I estimates from the previous step. Note that all the parameters were iterated upon convergence.
The derived stellar parameters for all 16 red giants are summarized in Table \ref{tab:Stars_param}.

\begin{table*}
\begin{center}
\caption{Stellar parameters.}\label{tab:Stars_param}
{\small
 \begin{tabular}{cccccccc}
\hline
 \raisebox{-1.5ex}{Star ID} & T$_{\rm eff}^{\rm 2MASS}$ & T$_{\rm eff}^{(V-I)}$   & T$_{\rm eff}^{\rm spec}$ & $\log$~g & $\xi$ & [Fe I/H]  & [Fe II/H] \\
         & [K]         &   [K]     &   [K]    &   $\log\mathrm{[cm\,s^{-2}]}$ &  [\kms]   & [dex]         &     [dex]      \\
\hline
  239   &   4021  &  4208  & 4300   &  1.07  &  2.1   &   $-1.19 \pm   0.03$	& $-1.17  \pm 0.06$ \\
  251   &   3741  &  3958  & 3900   &  0.49  &  2.2   &   $-1.08 \pm   0.04$	& $-0.76  \pm 0.05$ \\
  442   &   3914  &  4152  & 4100   &  0.82  &  2.1   &   $-1.13 \pm   0.03$	& $-0.94  \pm 0.08$ \\
  461   &   3686  &  3938  & 3800   &  0.31  &  2.1   &   $-1.03 \pm   0.04$	& $-0.66  \pm 0.06$ \\
  483   &   3928  &  4162  & 4150   &  0.87  &  2.0   &   $-1.10 \pm   0.03$	& $-1.02  \pm 0.09$ \\
  486   &   4082  &  4308  & 4400   &  1.23  &  2.1   &   $-1.09 \pm   0.03$	& $-1.02  \pm 0.06$ \\
  503   &   3702  &  3932  & 3850   &  0.36  &  2.3   &   $-1.24 \pm   0.03$	& $-0.86  \pm 0.04$ \\
  512   &   4060  &  4232  & 4250   &  1.07  &  2.0   &   $-1.19 \pm   0.03$	& $-1.04  \pm 0.02$ \\
  583\tablefootmark{1}   &   3747  &  3901  & 3950   &  0.42  &  2.2   &   $-1.24 \pm   0.04$	& $-1.11  \pm 0.06$ \\
  612\tablefootmark{1}   &   4231  &  4379  & 4400   &  1.25  &  2.1   &   $-1.18 \pm   0.02$	& $-0.90  \pm 0.03$ \\
  655   &   3954  &  4152  & 4150   &  0.87  &  2.1   &   $-1.21 \pm   0.03$	& $-1.08  \pm 0.05$ \\
  876   &   4144  &  4322  & 4350   &  1.26  &  2.0   &   $-1.13 \pm   0.03$	& $-1.01  \pm 0.03$ \\
  901   &   3872  &  4091  & 4100   &  0.78  &  2.1   &   $-1.11 \pm   0.03$	& $-0.96  \pm 0.04$ \\
 1251   &   4088  &  4232  & 4240   &  1.05  &  2.4   &   $-1.23 \pm   0.04$	& $-1.04  \pm 0.09$ \\
 1312   &   3763  &  3980  & 3950   &  0.53  &  2.4   &   $-1.31 \pm   0.04$	& $-1.14  \pm 0.07$ \\
 1459\tablefootmark{1}   &   3915  &  4130  & 4150   &  0.81  &  2.1   &   $-1.17 \pm   0.03$	& $-1.02  \pm 0.04$ \\
\hline
 \end{tabular}
\par}
\tablefoot{
\tablefoottext{1}{Possible AGB stars.}
}
\end{center}
\end{table*}

\subsection{Abundance errors}

We calculated the random error of the abundances as $\sigma_{EW} / \sqrt{N}$, where $N$ is the number of lines used for those elements where we measured more than one line and $\sigma_{EW}$ is the standard deviation. For those elements, for which we used the EW of a single line, we adopted a typical random error based on the mean abundance spread of all stars in our sample. For the abundances derived via synthesis, we adopted random errors based on the minimum and maximum abundance values that still yielded acceptable fits to the observed spectra.

To investigate the systematic abundance errors caused by the uncertainties of the stellar atmosphere parameters, we calculated a grid of new model atmospheres, varying the effective temperature by $\pm60$~K, the surface gravity by $\pm0.2$~dex, the microturbulent velocity by $\pm0.1$~\kms and the metallicity by $\pm0.1$~dex for three stars spanning a large difference in their parameters.
We also included calculations for Solar [$\alpha$/Fe] ratios using the Kurucz ODFNEW models (column labeled 'ODF' in Table \ref{tab:sys_errors}). Then we recomputed the abundances of all elements with the modified atmosphere models. The results are summarized in Table \ref{tab:sys_errors} in terms of difference to the default values.
The column labeled {\it total} lists the errors of all parameters combined in quadrature, including a $0.1$~dex uncertainty in [$\alpha$/Fe]. The latter corresponds to 1/4 of the abundance difference between the AODFNEW and ODFNEW Kurucz models. We note, however, that these are upper limits due to the covariance of the atmospheric parameters \citep[e.g.][]{mcwilliam+95}.

The change in temperature has a larger effect on the species with a lower excitation potential (e.g. K I, Ti I, V I, Cr I), whilst the change in gravity affects mostly the ionised species. For warm GK giants the dominant Fe II species are more sensitive to gravity and the Fe I species to changes in temperature. But we note that the stars studied here, especially the coolest stars of our sample, have so low temperatures that they are in the transition from Fe being dominated by the ionised species to Fe dominated by the neutral species. That is why Fe II is so sensitive to variations of the effective temperature in this case.
We note that our prior ignorance of the metallicity of the model atmospheres has only a negligible effect on the derived abundances.
The overall, typical, systematic uncertainties are of the order of $0.1$~dex.

\begin{table*}
\begin{center}
\caption{Systematic abundance errors.}\label{tab:sys_errors}
{\footnotesize
 \begin{tabular}{ccccccccccc}
\hline
 Ion & \multicolumn{2}{c}{$\Delta T_{eff}$} & \multicolumn{2}{c}{$\Delta\log g$}  & \multicolumn{2}{c}{$\Delta\xi$} & \multicolumn{2}{c}{$\Delta$[M/H]}  & ODF  & total \\

         & $+$60~K & $-$60~K & $+$0.2~dex & $-$0.2~dex & $+$0.1~dex  & $-$0.1~dex & $+$0.1~km/s & $-$0.1~dex & & \\
\hline
\multicolumn{11}{c}{\#461}\\
\hline
Fe I  & $-$0.02 &$+$0.01 &     $+$0.02& $-$0.06   &   $-$0.08& $+$0.06&     $+$0.03& $-$0.02   &    $-$0.12   &  0.09\\
Fe II & $-$0.14 &$+$0.14 &     $+$0.06& $-$0.13   &   $-$0.03& $+$0.03&     $+$0.08& $-$0.00   &    $-$0.18   &  0.18\\
O I   & $+$0.02 &$-$0.02 &     $+$0.07& $-$0.08   &   $-$0.01& $+$0.01&     $+$0.05& $-$0.03   &    $-$0.15   &  0.10\\
Na I  & $+$0.06 &$-$0.05 &     $-$0.02& $+$0.01   &   $-$0.01& $+$0.02&     $ $0.00& $+$0.01   &    $-$0.01   &  0.06\\
Mg I  & $-$0.01 &$+$0.02 &     $ $0.00& $-$0.03   &   $-$0.04& $+$0.05&     $+$0.02& $+$0.01   &    $-$0.05   &  0.05\\
Al I  & $+$0.04 &$-$0.04 &     $-$0.01& $ $0.00   &   $-$0.01& $+$0.01&     $ $0.00& $ $0.00   &    $-$0.02   &  0.04\\
Si I  & $-$0.07 &$+$0.08 &     $+$0.02& $-$0.06   &   $-$0.01& $+$0.02&     $+$0.05& $+$0.01   &    $-$0.09   &  0.09\\
K I   & $+$0.06 &$-$0.06 &     $ $0.00& $-$0.02   &   $-$0.10& $+$0.09&     $+$0.02& $-$0.01   &    $-$0.18   &  0.12\\
Ca I  & $+$0.07 &$-$0.05 &     $-$0.01& $+$0.01   &   $-$0.06& $+$0.08&     $+$0.02& $+$0.01   &    $-$0.08   &  0.10\\
Sc II & $-$0.01 &$+$0.02 &     $+$0.06& $-$0.08   &   $-$0.03& $+$0.05&     $+$0.06& $-$0.01   &    $-$0.13   &  0.09\\
Ti I  & $+$0.09 &$-$0.07 &     $+$0.02& $-$0.01   &   $-$0.07& $+$0.08&     $+$0.02& $-$0.01   &    $-$0.13   &  0.12\\
Ti II & $-$0.03 &$+$0.02 &     $+$0.04& $-$0.05   &   $-$0.05& $+$0.04&     $+$0.05& $-$0.01   &    $-$0.12   &  0.08\\
V I   & $+$0.09 &$-$0.09 &     $+$0.02& $-$0.02   &   $-$0.07& $+$0.06&     $+$0.03& $-$0.03   &    $-$0.14   &  0.12\\
Cr I  & $+$0.07 &$-$0.07 &     $ $0.00& $-$0.01   &   $-$0.08& $+$0.07&     $+$0.01& $-$0.01   &    $-$0.10   &  0.11\\
Mn I  & $+$0.02 &$ $0.00 &     $+$0.02& $-$0.04   &   $-$0.07& $+$0.08&     $+$0.04& $ $0.00   &    $-$0.09   &  0.09\\
Co I  & $+$0.01 &$ $0.00 &     $+$0.04& $-$0.05   &   $-$0.07& $+$0.09&     $+$0.05& $-$0.01   &    $-$0.11   &  0.10\\
Ni I  & $+$0.01 &$-$0.01 &     $+$0.04& $-$0.06   &   $-$0.05& $+$0.05&     $+$0.04& $-$0.01   &    $-$0.11   &  0.08\\
Cu I  & $+$0.01 &$ $0.00 &     $+$0.04& $-$0.06   &   $-$0.08& $+$0.09&     $+$0.05& $-$0.01   &    $-$0.12   &  0.11\\
Zn I  & $-$0.07 &$+$0.07 &     $-$0.01& $-$0.03   &   $-$0.04& $+$0.04&     $+$0.03& $+$0.02   &    $-$0.05   &  0.09\\
Y II  & $ $0.00 &$ $0.00 &     $+$0.05& $-$0.07   &   $-$0.04& $+$0.04&     $+$0.04& $-$0.02   &    $-$0.13   &  0.08\\
Ba II & $+$0.01 &$ $0.00 &     $+$0.06& $-$0.07   &   $-$0.10& $+$0.12&     $+$0.06& $-$0.02   &    $-$0.17   &  0.14\\
La II & $+$0.02 &$-$0.02 &     $+$0.06& $-$0.07   &   $-$0.03& $+$0.04&     $+$0.05& $-$0.02   &    $-$0.13   &  0.09\\
Ce II & $+$0.01 &$ $0.00 &     $+$0.06& $-$0.07   &   $-$0.01& $+$0.02&     $+$0.04& $-$0.02   &    $-$0.12   &  0.08\\
Pr II & $+$0.02 &$-$0.02 &     $+$0.03& $-$0.07   &   $-$0.03& $+$0.02&     $+$0.04& $-$0.03   &    $-$0.13   &  0.08\\
Nd II & $+$0.01 &$-$0.01 &     $+$0.05& $-$0.06   &   $-$0.06& $+$0.06&     $+$0.05& $-$0.02   &    $-$0.13   &  0.09\\
Sm II & $+$0.02 &$-$0.02 &     $+$0.05& $-$0.05   &   $-$0.05& $+$0.06&     $+$0.04& $-$0.01   &    $-$0.12   &  0.09\\
Eu II & $-$0.01 &$+$0.01 &     $+$0.06& $-$0.08   &   $-$0.02& $+$0.02&     $+$0.05& $-$0.02   &    $-$0.14   &  0.09\\
\multicolumn{11}{c}{\#612}\\
\hline
Fe I  & $+$0.05& $-$0.05&      $+$0.01& $-$0.03&      $-$0.06& $+$0.05&      $ $0.00& $-$0.01&       $-$0.04&	  0.08\\
Fe II & $-$0.08& $+$0.08&      $+$0.08& $-$0.12&      $-$0.03& $+$0.03&      $+$0.03& $-$0.04&       $-$0.14&	  0.14\\
Na I  & $+$0.04& $-$0.05&      $-$0.01& $+$0.01&      $-$0.02& $+$0.01&      $-$0.01& $ $0.00&       $+$0.02&	  0.05\\
Mg I  & $+$0.04& $-$0.04&      $-$0.01& $ $0.00&      $-$0.03& $+$0.03&      $ $0.00& $+$0.01&       $ $0.00&	  0.05\\
Al I  & $+$0.04& $-$0.05&      $-$0.01& $ $0.00&      $-$0.01& $ $0.00&      $-$0.01& $ $0.00&       $+$0.01&	  0.05\\
Si I  & $-$0.02& $+$0.03&      $+$0.03& $-$0.04&      $-$0.01& $+$0.02&      $+$0.02& $-$0.01&       $-$0.05&	  0.05\\
K I   & $+$0.11& $-$0.10&      $+$0.01& $-$0.01&      $-$0.06& $+$0.07&      $-$0.01& $+$0.02&       $-$0.04&	  0.13\\
Ca I  & $+$0.06& $-$0.07&      $-$0.02& $ $0.00&      $-$0.05& $+$0.04&      $-$0.02& $+$0.01&       $ $0.00&	  0.08\\
Sc II & $-$0.02& $+$0.01&      $+$0.06& $-$0.10&      $-$0.04& $+$0.03&      $+$0.03& $-$0.03&       $-$0.11&	  0.10\\
Ti I  & $+$0.12& $-$0.13&      $ $0.00& $ $0.00&      $-$0.04& $+$0.04&      $-$0.01& $+$0.01&       $-$0.01&	  0.13\\
Ti II & $-$0.02& $+$0.02&      $+$0.06& $-$0.09&      $-$0.05& $+$0.05&      $+$0.02& $-$0.03&       $-$0.11&	  0.10\\
V I   & $+$0.11& $-$0.12&      $+$0.01& $+$0.01&      $-$0.01& $+$0.01&      $ $0.00& $+$0.01&       $ $0.00&	  0.12\\
Cr I  & $+$0.11& $-$0.11&      $ $0.00& $+$0.01&      $-$0.05& $+$0.06&      $-$0.01& $+$0.02&       $ $0.00&	  0.12\\
Mn I  & $+$0.09& $-$0.09&      $+$0.01& $-$0.01&      $-$0.05& $+$0.06&      $ $0.00& $+$0.01&       $-$0.02&	  0.11\\
Co I  & $+$0.07& $-$0.05&      $+$0.03& $-$0.02&      $-$0.02& $+$0.04&      $+$0.02& $ $0.00&       $-$0.03&	  0.07\\
Ni I  & $+$0.04& $-$0.04&      $+$0.03& $-$0.04&      $-$0.03& $+$0.03&      $+$0.01& $-$0.01&       $-$0.05&	  0.06\\
Cu I  & $+$0.06& $-$0.05&      $+$0.03& $-$0.03&      $-$0.06& $+$0.07&      $+$0.01& $-$0.01&       $-$0.04&	  0.09\\
Zn I  & $-$0.04& $+$0.04&      $+$0.04& $-$0.07&      $-$0.03& $+$0.03&      $+$0.02& $-$0.02&       $-$0.07&	  0.08\\
Y II  & $-$0.01& $+$0.01&      $+$0.07& $-$0.08&      $-$0.03& $+$0.05&      $+$0.03& $-$0.03&       $-$0.10&	  0.09\\
Ba II & $+$0.02& $-$0.02&      $+$0.07& $-$0.09&      $-$0.08& $+$0.09&      $+$0.03& $-$0.03&       $-$0.14&	  0.13\\
La II & $+$0.01& $-$0.01&      $+$0.08& $-$0.08&      $-$0.01& $+$0.02&      $+$0.04& $-$0.03&       $-$0.11&	  0.09\\
Ce II & $ $0.00& $ $0.00&      $+$0.08& $-$0.08&      $-$0.01& $+$0.01&      $+$0.03& $-$0.03&       $-$0.10&	  0.09\\
Pr II & $+$0.01& $-$0.02&      $+$0.07& $-$0.09&      $-$0.01& $+$0.01&      $+$0.03& $-$0.04&       $-$0.11&	  0.09\\
Nd II & $ $0.00& $-$0.01&      $+$0.07& $-$0.09&      $-$0.03& $+$0.03&      $+$0.03& $-$0.04&       $-$0.11&	  0.10\\
Sm II & $+$0.01& $-$0.01&      $+$0.08& $-$0.08&      $-$0.01& $+$0.02&      $+$0.04& $-$0.03&       $-$0.09&	  0.09\\
Eu II & $-$0.02& $+$0.01&      $+$0.07& $-$0.10&      $-$0.01& $+$0.01&      $+$0.03& $-$0.04&       $-$0.12&	  0.10\\
\hline
 \end{tabular}

\par}
\end{center}
\end{table*}

\section{Abundance results}

\begin{figure}
\centering
\resizebox{\hsize}{!}{
\includegraphics[angle=0]{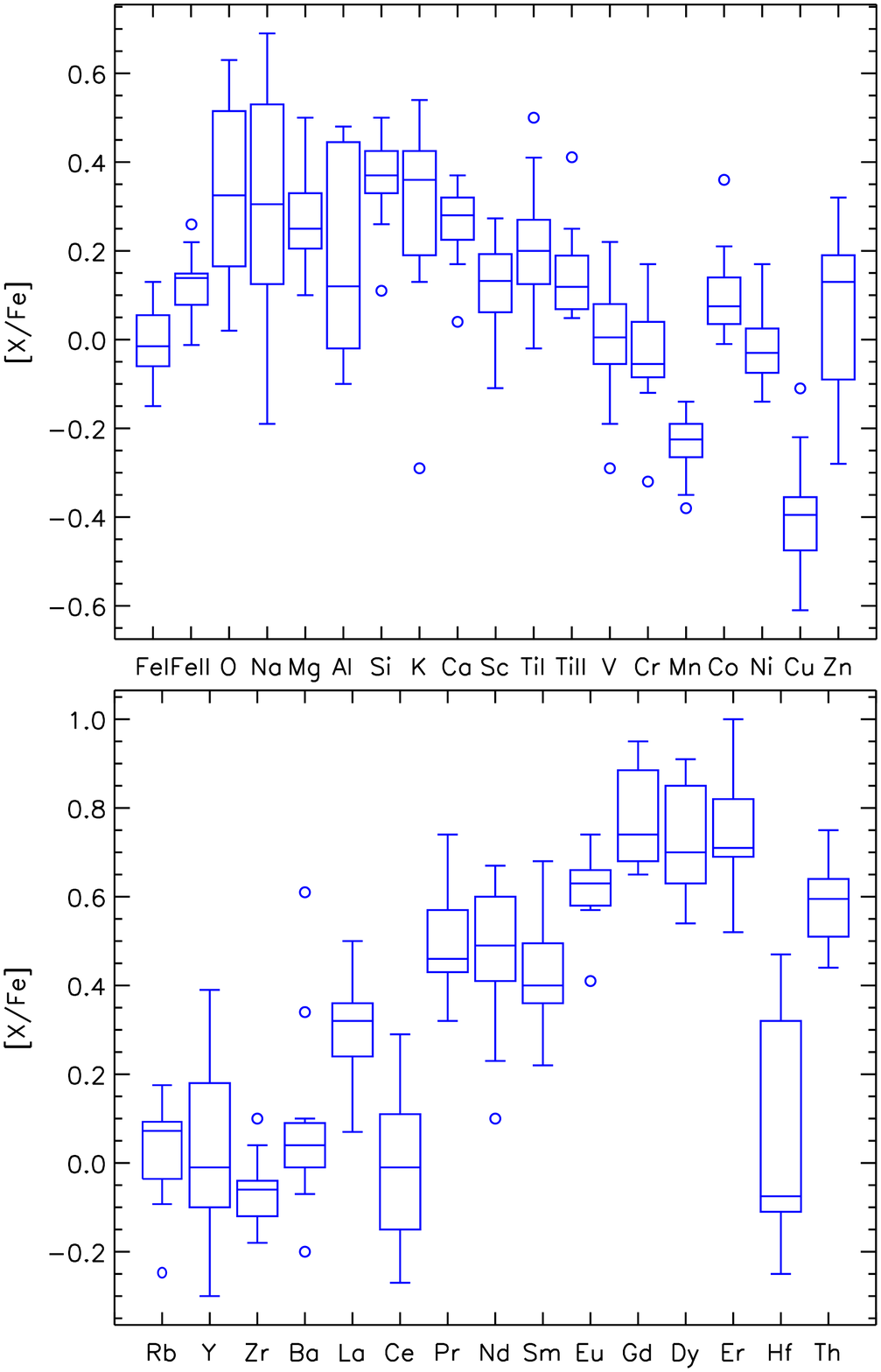}
}
\caption[]{Boxplot of the derived abundances in M~75 relative to iron. Fe I and Fe II abundances are relative to the Fe I cluster mean. Neutral species are relative to Fe I and ionised species are relative to Fe II. The boxes designate the median values and IQR. The error bars show the minimum and the maximum value. Outliers are shown with circles. An outlier is defined if it deviates by more than $1.5$~times the IQR.
}
\label{fig:iqr}
\end{figure}

In Table \ref{tab:spreads} we summarize the abundance results for M~75, relative to Fe I for all neutral species and to Fe II for all ionised species. We also list the mean random error, $\epsilon_{rand}$, and the mean systematic error, $\epsilon_{sys}$, on the abundance ratios [X/Fe]. The columns labeled $\sigma_{obs}^{16}$ and $\sigma_{obs}^{13}$ contain the observed spreads of the abundances within the cluster.
As noted above, the discrepancy between Fe I and Fe II values are largest for the coolest stars in our sample, leading to a larger spread in the Fe II abundance. For this reason, we show the spreads calculated by using all stars ($\sigma_{obs}^{16}$) and by excluding the three coolest ones ($\sigma_{obs}^{13}$).
The last two columns, $\sigma_0^{16}$ and $\sigma_0^{13}$, show the cluster's intrinsic spreads for all stars and without the three coolest stars, respectively, obtained by correcting for the measurement uncertainty as:
\begin{equation}
\sigma_0^2 = \sigma_{obs}^2 - \epsilon_{rand}^2.
\end{equation}
Note that Equation 2 gives an over-estimate of the intrinsic dispersion, because the true systematic uncertainties were not removed. 
Figure \ref{fig:iqr} shows the interquartile ranges (IQR) and the median values of the abundances we derived.
The only siginificant intrinsic spreads were found for the light elements O, Na, Al, and the s-process element Ba. We also note the presence of one K-deficient star, which, however, does not present any anomalous O, Mg, Na, or Al abundances. The scatters of all other elements are compatible with the observational errors.

A table containing all chemical element abundance ratios with associated random errors for all individual stars is available in the electronic version of the journal. A part of it is presented in Table \ref{tab:Stars_abund} to guide the eye. The column labeled N shows the number of lines used to derive the particular element abundance and $\epsilon_{rand}$ shows the random error.

\begin{table*}
\begin{center}
\caption{Derived abundance ratios for the individual stars of the GC. The full table is available in the online version of the journal.}\label{tab:Stars_abund}
{\footnotesize
 \begin{tabular}{cccccccccccccccccc}
\hline
Star ID & [FeI/H] & $\epsilon_{rand}$ & N & [FeII/H] & $\epsilon_{rand}$ & N & [O/H] & $\epsilon_{rand}$ & N & [Na/H] & $\epsilon_{rand}$ & N & [Mg/H] & $\epsilon_{rand}$ & N & ... \\
\hline
 239  &    $-$1.19 &   0.03 & 66 &  $-$1.17  &    0.06 & 6 & $-$0.62  &      0.01 & 2 &$-$1.20 &   0.08 &3& $-$0.97  &  0.1 & 1&   \\
 251  &    $-$1.08 &   0.04 & 46 &  $-$0.76  &    0.05 & 5 & $-$0.81  &      0.01 & 2 &$-$0.87 &   0.11 &3& $-$0.66  &  0.1 & 1&   \\
 442  &    $-$1.13 &   0.03 & 60 &  $-$0.94  &    0.08 & 4 & $-$0.95  &      0.12 & 2 &$-$0.62 &   0.10 &3& $-$0.83  &  0.1 & 1&   \\
 461  &    $-$1.03 &   0.04 & 42 &  $-$0.66  &    0.06 & 3 & $-$0.51  &      0.04 & 2 &$-$1.22 &   0.04 &3& $-$0.93  &  0.1 & 1&   \\
 483  &    $-$1.10 &   0.03 & 64 &  $-$1.02  &    0.09 & 5 & $-$0.99  &      0.04 & 2 &$-$0.41 &   0.06 &3& $-$0.79  &  0.1 & 1&   \\
 486  &    $-$1.09 &   0.03 & 68 &  $-$1.02  &    0.06 & 5 & $-$1.07  &      0.04 & 2 &$-$0.70 &   0.02 &3& $-$0.76  &  0.1 & 1&   \\
 503  &    $-$1.24 &   0.03 & 49 &  $-$0.86  &    0.04 & 6 & $-$0.88  &      0.08 & 2 &$-$0.95 &   0.05 &3& $-$0.88  &  0.1 & 1&   \\
 512  &    $-$1.19 &   0.03 & 66 &  $-$1.04  &    0.02 & 5 & $-$0.56  &      0.11 & 2 &$-$1.23 &   0.11 &3& $-$1.00  &  0.1 & 1&   \\
 583  &    $-$1.24 &   0.04 & 52 &  $-$1.11  &    0.06 & 4 & $-$0.79  &      0.18 & 2 &$-$1.20 &   0.06 &3& $-$1.01  &  0.1 & 1&   \\
 612  &    $-$1.18 &   0.02 & 66 &  $-$0.90  &    0.03 & 5 & $-$0.89  &      0.15 & 0 &$-$0.90 &   0.06 &3& $-$1.04  &  0.1 & 1&   \\
 655  &    $-$1.21 &   0.03 & 63 &  $-$1.08  &    0.05 & 7 & $-$0.76  &      0.02 & 2 &$-$0.89 &   0.08 &3& $-$0.88  &  0.1 & 1&   \\
 876  &    $-$1.13 &   0.03 & 63 &  $-$1.01  &    0.03 & 5 & $-$0.62  &      0.08 & 2 &$-$0.91 &   0.09 &3& $-$0.90  &  0.1 & 1&   \\
 901  &    $-$1.11 &   0.03 & 57 &  $-$0.96  &    0.04 & 5 & $-$1.01  &      0.04 & 2 &$-$0.49 &   0.10 &3& $-$0.87  &  0.1 & 1&   \\
1251  &    $-$1.23 &   0.04 & 54 &  $-$1.04  &    0.09 & 4 & $-$1.02  &      0.1  & 1 &$-$0.61 &   0.06 &3& $-$0.73  &  0.1 & 1&   \\
1312  &    $-$1.31 &   0.04 & 51 &  $-$1.14  &    0.07 & 6 & $-$0.70  &      0.06 & 2 &$-$0.95 &   0.07 &3& $-$1.05  &  0.1 & 1&   \\
1459  &    $-$1.17 &   0.03 & 61 &  $-$1.02  &    0.04 & 7 & $-$1.02  &      0.1  & 1 &$-$0.62 &   0.13 &3& $-$1.04  &  0.1 & 1&   \\
\hline
 \end{tabular}
\par}
\end{center}
\end{table*}

\begin{table}
\begin{center}
\caption{Average abundance ratios, average random and systematic errors, observational, and intrinsic spreads. See text for details.}\label{tab:spreads}
{\footnotesize
 \begin{tabular}{crcccccc}
\hline
 Element & [X/Fe] & $\epsilon_{rand}$ &  $\epsilon_{sys}$ & $\sigma_{obs}^{16}$ & $\sigma_{obs}^{13}$  & $\sigma_0^{16}$ & $\sigma_0^{13}$  \\
\hline
   Fe I/H  &   $-$1.16  &	 0.03	& 0.08 & 0.07   &   0.06   &  0.06  &	 0.05 \\   
  Fe II/H  &   $-$1.03  &	 0.05	& 0.16 & 0.13   &   0.07   &  0.12  &	 0.04 \\   
     O I &    0.34  &	 0.08	& 0.10 &  0.20   &   0.21 &  0.18  &	0.19	 \\        
    Na I &    0.30  &	 0.08	& 0.06 &  0.25   &   0.24    &  0.24  &    0.22     \\     
    Mg I &    0.27  &	 0.11	& 0.05 &  0.10   &   0.09    &  0.00  &    0.00     \\     
    Al I &    0.19  &	 0.08	& 0.05 &  0.22   &   0.22    &  0.20  &    0.20     \\     
    Si I  &    0.37  &	 0.07	& 0.07 &  0.09   &   0.09    &  0.06  &    0.04     \\     
     K I &    0.30  &	 0.09	& 0.13 &  0.20   &   0.11    &  0.18  &    0.06     \\     
    Ca I &    0.26  &	 0.06	& 0.09 &  0.08   &   0.05    &  0.05  &    0.00     \\     
    Sc II &    0.08  &	 0.12	& 0.09 &  0.13   &   0.10    &  0.05  &    0.00     \\     
   Ti I  &    0.21  &	 0.07	& 0.13 &  0.13   &   0.14    &  0.11  &    0.12     \\     
  Ti II  &    0.09  &	 0.10	& 0.09 &  0.15   &   0.10    &  0.11  &    0.00     \\     
     V I &    0.00  &	 0.05	& 0.12 &  0.12   &   0.10    &  0.11  &    0.09     \\     
    Cr I &   $-$0.03  &	 0.11	& 0.12 &  0.11   &   0.09   &  0.00  &    0.00     \\ 	   
    Mn I &   $-$0.24  &	 0.06	& 0.09 &  0.06   &   0.05    &  0.00  &    0.00     \\     
    Co I &    0.10  &	 0.06	& 0.08 &  0.09   &   0.09    &  0.07  &    0.07     \\     
    Ni I &   $-$0.02  &	 0.06	& 0.07 &  0.08   &   0.08    &  0.05  &    0.05     \\     
    Cu I &   $-$0.39  &	 0.13	& 0.10 &  0.11   &   0.10    &  0.00  &    0.00     \\     
    Zn I &    0.05  &	 0.15	& 0.08 &  0.17   &   0.16    &  0.08  &    0.06     \\     
    Rb I &    0.02  &	 0.15	& 0.09 &  0.15   &   0.07    &  0.00  &    0.00     \\     
     Y II &   0.03  &    0.21	& 0.10 &  0.21   &   0.19  &	0.00 &    0.00   \\	   
    Zr II &   $-$0.06  & 0.16	& 0.08 &  0.15   &   0.07  &	0.00 &    0.00   \\	   
    Ba II &    0.08  &	 0.07	& 0.13 &  0.22   &   0.19    &    0.21  &    0.18   \\     
    La II &    0.31  &	 0.12	& 0.09 &  0.18   &   0.11    &    0.13  &    0.00   \\     
    Ce II &    0.00  & 0.16	& 0.08 &  0.22   &   0.18   &	 0.15  &    0.08   \\ 	   
    Pr II &    0.50  &	 0.16	& 0.09 &  0.15   &   0.12    &    0.00  &    0.00   \\     
    Nd II &    0.46  &	 0.16	& 0.09 &  0.18   &   0.17    &    0.08  &    0.06   \\     
    Sm II &    0.42  &	 0.09	& 0.09 &  0.12   &   0.11    &    0.08  &    0.08   \\     
    Eu II &    0.62  &	 0.07	& 0.09 &  0.14   &   0.08    &    0.12  &    0.04   \\     
    Gd II &    0.78  &	 0.21	& 0.10 &  0.11   &   0.11    &    0.00  &    0.00   \\     
    Dy II &    0.74  &	 0.21	& 0.10 &  0.17   &   0.12    &    0.00  &    0.00   \\	   
    Er II &    0.74  &	 0.21	& 0.10 &  0.20   &   0.12    &    0.00  &    0.00   \\	   
    Hf II &    0.06  &	 0.21	& 0.10 &  0.27   &   0.26    &    0.17  &    0.15   \\	   
    Th II &    0.59  &	 0.21	& 0.10 &  0.19   &   0.09    &    0.00  &    0.00   \\	   
\hline 
 \end{tabular}
\par}
\end{center}
\end{table}


\subsection{Iron}

With this study we derived the first measurement of the Fe abundance of M~75 based on high-resolution spectroscopy as [Fe I/H]~$=-1.16 \pm 0.03$~dex (random) $\pm0.08$~dex (systematic) with a marginal $1\sigma$ spread of $0.07$~dex.
Using Fe II lines, we obtained a higher value of [Fe II/H]~$=-0.98\pm0.03$~dex (random) $\pm0.16$~dex (systematic) with a $1\sigma$ spread of $0.13$~dex. There is no trend of the [Fe I/H] or [Fe II/H] values with temperature, except for the three coolest stars, which have considerably higher [Fe II/H]. Excluding these three coolest stars, the mean discrepancy between the Fe I and Fe II values becomes [Fe I/Fe II]~$=-0.14\pm0.02$~dex, which is still significant but the larger scatter of Fe II abundances is reduced to the same value as the Fe I scatter ($\sigma = 0.07$~dex). 

Both values are in an excellent agreement with the metallicity derived by \citet{catelan+2002}, based on UBV photometry of [Fe/H]~$=-1.03\pm0.17$~dex and [Fe/H]~$=-1.24\pm0.21$~dex on the metallicity scales of \citet{carretta+gratton1997} and \citet{zinn+west1984}, respectively.

The scatter of $0.07$~dex may seem large compared to the bulk of GCs \citep[$\sigma \lesssim 0.05$~dex;][]{carretta+2009} but it is fully consistent with the higher luminosity of M~75. With an absolute magnitude of M$_{\rm V} = -8.57$~mag \citep{harris96}, it is amongst the most luminous and hence most massive GCs in the Milky Way (only $18$ of the $\sim150$ MW GCs are brighter). \citet{carretta+2009} reported a dependence on the scatter in [Fe/H] with various cluster parameters. In their homogenous sample of 19 GCs they showed that the $1\sigma$ dispersion correlates with the cluster luminosity (a proxy for the present mass), the maximum effective temperature reached on the HB, and anticorrelates with the level of $\alpha$-enrichment.
All these correlations point to a better capability of more massive clusters to self-enrich with the ejecta of massive stars. The deeper gravitational potential helps retaining the massive stars ejecta and the hotter HB stars indicate larger He-content and thus, more processing by previous generation of stars. Our sample of 16 stars, however, does not allow us to make any firm conclusions on the link between the observed Fe-scatter in M~75 and its self-enrichment.


\subsection{Alpha elements}

The production of $\alpha$-elements such as O, Ne, Mg, Si, S, Ca, and Ti is mainly associated with the eruptions of SNe II. The different timescales of the occurrence of SNe II and SN Ia make the [$\alpha$/Fe] abundance ratios a powerful tool for diagnosing the chemical evolution and star formation history (SFH) of any stellar population \citep{tinsley1979}.
In the Milky Way, the relatively metal poor halo stars form a plateau of enhanced [$\alpha\rm{/Fe]}\sim+0.4$~dex, which starts to decrease when [Fe/H]$~\gtrsim -1.0$~dex due to the onset of SNe Ia.
A high [$\alpha$/Fe] ratio is associated with rapid star formation episodes that ceased before the long-lived SNe Ia, the main source of iron, began to enrich the local environment through their ejecta. Dwarf galaxies, on the other hand, have slower star formation rates so that low values of [$\alpha/$Fe] are observed already at low metallicities \citep[e.g.,][]{shetrone+2001,shetrone+2003,tolstoy+09}. Although different $\alpha$-elements are produced on similar timescales, they show element-to-element variations due to different production mechanisms, either through hydrostatic He-burning in the cores of massive stars (e.g., O and Mg), or during the SNe explosions themselves (e.g., Si, Ca, and Ti).

In M~75, we derived O-abundances by measuring the EWs and by spectral synthesis of the $6300$~\AA~and $6364$~\AA~lines, which are free of telluric contamination owing to the fortunate radial velocity of this GC. The $6364$~\AA~line is, however, situated in the wing of a broad Ca autoionisation feature, so in the synthesis of this line we adopted the derived Ca-abundance. Additionally, we adopted Solar C- and N-abundances but we confirmed the results from \citet{koch+2009} that demonstrated that molecular (CNO) equilibrium does not affect the derived O-abundances at these metallicities and levels of O-enhancements. We also confirmed that there are not any extreme variations in the strength of the CH G-band around $4320$~\AA.

Mg-abundances were derived from the line at $5711$~\AA, Si from the $5684$~\AA, $5949$~\AA, and $6155$~\AA~lines, Ca-abundances were based on 11 features between $5250$~\AA~and~$6750$~\AA, and Ti was measured from various Ti I and Ti II absorption lines. 
We found that all $\alpha$-elements are enhanced with respect to the Sun to a different extent. The average [$\alpha$/Fe] ratio is $0.3\pm0.02$~dex, based on the Mg, Si, and Ca abundances. This is consistent with the canonical value for the old stellar population of the Milky Way \citep[halo field stars and the majority of its GCs;][]{pritzl+2005}. Oxygen is the only $\alpha$-element, which shows significant variations in its abundance amongst the stars in our sample. These variations are discussed in the following section in terms of multiple populations.



\subsection{Proton-capture elements}

Elements like Na, Al, and K are produced through proton-capture reactions at high temperatures during the H-burning in the cores of massive stars. The above are the elements responsible for creating the unique chemical pattern of GCs \citep[e.g.][]{denisenkov+denisenkova89,langer+93}, also see the review from \citet{gratton+2004}. Large variations in the abundances of Na and Al have been so far detected in all GCs studied to date \citep{gratton+2012}. Nowadays, it is largely accepted that these variations are due to the presence of at least two stellar populations in every GC characterised by slightly different ages and abundance patterns.
Whilst both populations show the same content of Fe-peak elements, the later formed stars are characterised by enhanced N, Na, and Al, along with depleted C, O, and possibly Mg.
M~75 is not an exception in this respect. We measured Na-abundances from the three lines at $5682.6$, $6154.2$, and $6160.8$~\AA~and Al-abundances from the four lines at $6696, 6699, 7835$, and $7836$~\AA. Potassium abundances were derived mainly from the $7699$~\AA~line, which was free of telluric absorption, however, saturated in the majority of our stars. We also used the weaker $7665$~\AA~K line for reference, which was, however, strongly blended with telluric features.
\citet{lind+2011} raised attention to possible large NLTE effects in the derived Na abundances from some commonly used lines. According to this study, however, in the regime of our stars (bright cool giants) the NLTE corrections are expected to be small (in the order $0.05 - 0.1$~dex for the lines that we used) and  we ignored them.

\begin{figure}
\centering
\resizebox{\hsize}{!}{
\includegraphics[angle=0]{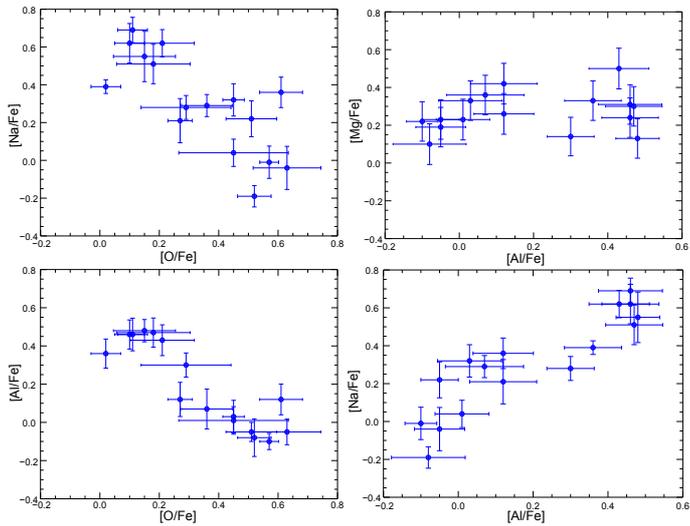}
}
\caption[]{Correlations between the light elements; upper left: Na-O anticorrelation; upper right: (no) Al-Mg anticorrelation; bottom left: O-Al anticorrelation; bottom right: Al-Na correlation.
}
\label{fig:anticor}
\end{figure}

Figure \ref{fig:anticor} shows different correlations between the light elements. The O-Na anticorrelation is clearly visible. We divided our sample into Primordial (P) population, characterised with O-rich and Na-poor stars, and Intermediate (I) population, characterised by O-poor and Na-rich stars, following the empirical separation introduced by \citet{carretta+2009b}.
The four stars with [Na/Fe]~$<0.1$~dex can be considered the remainder of the first generation that formed in the GC.
The number ratio of P- to I-generation stars is roughly $1/3$ based on the 16 stars in our sample, which is typical of most GCs but still rather low considering the cluster's high luminosity and the relatively large distance of our targets from the cluster's centre (median distance of $2.6$ half-light radii). 
\citet{carretta+2009b} reported a correlation between the ratio of P- to I-stars with the cluster's luminosity and with the median distance of the stars from the cluster's centre in agreement with the GC formation models by \citet{dercole+2008}, according to which, enriched by AGB stellar winds, gas is accumulated in the central regions of the cluster, where a kinematically cold second generation forms. Thus, in present-day GCs, the I-population is generally more concentrated in the central regions of a GC. Additionally, more massive clusters are better capable of retaining their stars, including the P-population. In this respect, we would expect to observe a P- to I-stars ratio of about $0.6$ in M~75 or in a sample of 16 outer stars -- 6 P-generation and 10 I-generation insted of the observed 4 P-generation and 12 I-generation stars.
However, the lower ratio of P- to I-stars in our sample could well be due to a low number statistics. It is also worth pointing out that there is a clear gap in the Na-O and Al-O anticorrelations, which might indicate the presence of more than two generations of stars in M~75 (Figure \ref{fig:anticor}). Assuming that the gap is real and not only owing to the small number statistics, we give possible 
explanations in the Discussion section.

M~75 has a negligible spread in Mg with respect to the observational errors and we do not see a correlation between Mg and Al, as found in other massive GCs \citep{gratton+2001,carretta+2009c}. This can be explained with the abscence of an Extreme (E) population in M~75, characterised by extremely low O, high Na, and lower Mg abundances. Although, a rough calculation shows that the Al enhancement of $+0.6$~dex could be obtained by reducing the Mg abundance by only $0.06$~dex, thus there is no need to be a strong slope in the [Mg/Fe] vs. [Al/Fe] diagram. The large Al variation is, however, strongly correlated with Na and anticorrelated with O (Figure \ref{fig:anticor}). Neither of these elements shows a trend with effective temperature and we can consider these correlations genuine properties of the cluster.
Potassium is not correlated with any of those elements and its marginal variations could be due to significant temperature-dependent NLTE effects in the strong $7699$~\AA~line \citep{zhang+2006}. Thus, the [K/Fe] ratios show a slight trend of decreasing abundance with decreasing temperature.



\subsection{Iron-peak elements}

All the odd-Z iron-peak elements V, Mn, Co, and Cu suffer from significant HFS corrections. These corrections vary from $0.2$ to $0.6$~dex for the different elements, with the largest effects for Cu. As a result, Mn and Cu are depleted with respect to iron by $0.24\pm0.06$~dex and $0.39\pm0.13$~dex, respectively. Such values are not unusual and are observed in a number of stellar systems \citep{cayrel+2004,mcwilliam+2003,mcwilliam+smecker-hane2005} and a number of Galactic GCs \citep[e.g.][]{koch+2009,koch+cote2010}. 
Cobalt is slightly overabundant with respect to iron with a value of [Co/Fe]~$=0.10\pm0.06$~dex. All even-Z elements (Cr, Ni, Zn) plus Sc and V trace the dominant iron production in the long-lived SNe Ia in that the [X/Fe] ratios are compatible with the solar values ([X/Fe]~$\sim0.0$~dex).

\subsection{Neutron-capture elements}

We derived chemical element abundance ratios for a large variety of n-capture elements mainly based on EW measurements, but we employed spectral synthesis for those elements, for which only few weak or highly blended lines were available. We used the Kurucz atomic database to obtain a blending list for our synthesis. The lighter n-capture elements Rb, Y, and Zr, usually associated with the weak s-process, have [X/Fe] ratios close to the Solar values (Table \ref{tab:spreads}).
We note, however, that the associated random errors on these ratios are large, owing to difficulties in measuring them.
For instance, Rb and Zr abundances are derived through a spectral synthesis of the $7800$~\AA~and $5112$~\AA~lines, respectively, which accounts for the severe blending of these lines.  
Yttrium abundances are derived based on EWs of three lines found in the blue region of the red arm of MIKE, which is generally characterised by a lower S/N ratio.
We also did not account for HFS effects associated with this odd-Z element due to lack of HFS data, but the corrections are expected to be small \citep[][]{prochaska+2000}.

Barium is the only n-capture element, which presents intrinsic variations significantly exceeding the random errors. In particular, there are two Ba-rich stars with [Ba/Fe]$=+0.34\pm0.10$~dex and [Ba/Fe]$=+0.61\pm0.06$~dex, whilst the mean [Ba/Fe] ratio is $0.01\pm0.02$~dex (Figure \ref{fig:BaFe}). Both stars are certainly not luminous and cool enough to have produced the s-process themselves and the high Ba abundances are most likely due to enrichment from AGB stars in the cluster's environment. The significance of these Ba-rich stars is further discussed in the Discussion section. Our Ba abundances were derived mostly based on EWs of the easily accessible $5854$~\AA~line, whilst the $6142$~\AA~and $6497$~\AA~lines are saturated in most of our stars. For the warmer stars with higher gravity, where the latter lines are not saturated (reduced EW $< -4.5$~dex), the derived abundances perfectly agree with the results from the $5854$~\AA~line alone.

Lanthanum, Ce, Pr, Nd, Sm, and Eu abundances were derived based on the EWs of one or two sufficiently strong lines, whilst the abundances for Gd, Dy, Er, Hf, and Th are derived based on spectral synthesis of one or several lines of these species. We note that the associated errors of the latter elements are large owing to the weakness of the lines and/or the low S/N ratios in their vicinity.
One might also be interested if the Ba-enhanced stars also show variations in the other s-process elements La and Ce. These two stars show a marginal enhancement in their La abundance by $\sim2\sigma$ above the cluster's average [La/Fe] ratio and statistically insignificant Ce-enhancement by less than $1\sigma$ above the cluster's average [Ce/Fe] ratio.
We applied HFS corrections to the odd-Z elements La and Eu but there were no available HFS data for Pr. We assumed that they are comparable in magnitude to the HFS corrections for the other n-capture elements and thus very small.

Besides Ba, Ce, and Hf, which show solar [X/Fe] ratios, all other heavy elements are enhanced with respect to the Sun; see Section 5.1 for discussion.

\section{Discussion}

\subsection{r- and s-process enrichment in M~75}

\begin{figure}
\centering
\resizebox{\hsize}{!}{
\includegraphics[angle=0]{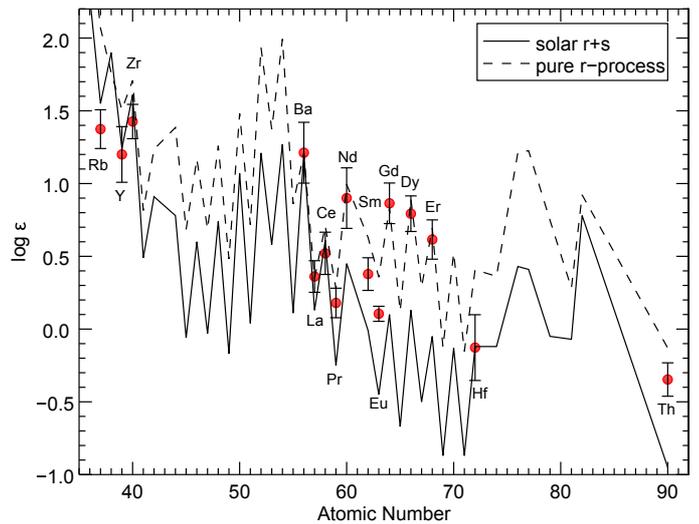}
}
\caption[]{Mean neutron capture element abundances for all stars in M 75, normalized to Ba. The lines display the scaled solar pure r- and r+s-process contributions from \citet{burris+2000}. The uncertainties represent the $1\sigma$ scatter of the derived element abundances in all stars.
}
\label{fig:r+s}
\end{figure}

\begin{figure}
\centering
\resizebox{\hsize}{!}{
\includegraphics[angle=0]{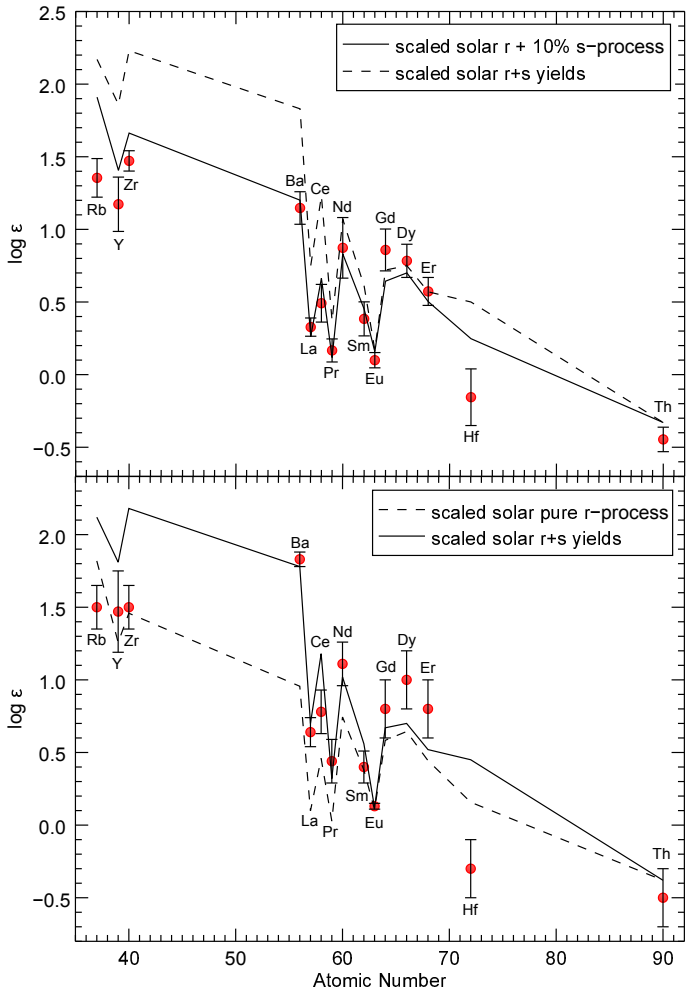}
}
\caption[]{Upper panel: Mean neutron-capture element abundance pattern for all ``Ba-normal'' (s-process deficient) stars in M~75. The solid line represents the best fit model for a scaled solar r-process plus $10\%$ of the scaled solar s-process yields. The dashed line shows the scaled solar total r+s-process yields for comparison. The error bars represent the $1\sigma$ scatter of the derived element abundances of all s-process deficient stars; Bottom panel: Derived n-capture elements abundances for star $\#901$ - the most s-process rich star in our sample. The solid line represents the best fit model for a scaled solar total r+s-process yields. The dashed line shows the scaled solar pure r-process model for comparison. The error bars represent the random errors for this particular star.  
}
\label{fig:r+s_fit}
\end{figure}

It is surprising that, at [Fe/H]~$=-1.16$~dex, M 75 seems to be one of the rarer cases of GCs compatible with predominant r-process production of the n-capture elements \citep[Figure \ref{fig:r+s}, see also][for M~15, M~5, and the distant GC Pal~3, respectively]{sneden+2000,yong+2008,koch+2009}. In Figure \ref{fig:r+s} we show the total r+s-process Solar curve and the pure r-process curve from \citet{burris+2000} plotted over the mean n-capture element abundances derived from our spectroscopic sample of M~75. Both curves are normalised to the mean Ba-abundance for an easier distinction.
A $\chi^2$ test shows that the best fit to the production of the elements from Ba (Z$=56$) to Th (Z$=90$) is found for a scaled Solar pure r-process enrichment plus an admixture of $10\%$ of the Solar s-process yields (Figure. \ref{fig:r+s_fit}; upper panel). In this fit we included all stars in our sample but the two Ba-enhanced ones and we excluded the elements Rb, Y, Zr, and Hf. Hafnium lies off the general pattern in all stars and we assumed that its abundances likely suffer from a severe systematic offset. The lighter n-capture elements Rb, Y, and Zr, on the other hand, have very complicated production channels, which are not yet fully understood \citep{travaglio+2004}. For instance they are associated with the weak s- and r-processes, which appear in massive (M~$\sim20M_{\odot}$) stars on similar timescales as the r-process production from SNe II \citep{raiteri+1993}.
We conclude that only a small number of AGB stars have contributed to the enrichment of the primordial cloud from which M~75 formed. 

We note that there is not any difference in the n-capture element abundance pattern between the P- and I-generations in our sample with the exception of the two Ba-enhanced stars. We found the best $\chi^2$-fit for both Ba-rich stars to be scaled solar r- plus an admixture of $50\%$ and $100\%$ of the scaled Solar s-process yields, respectively. The abundance pattern of the Ba-rich star $\#901$ together with its best fitting model is presented in Figure \ref{fig:r+s_fit}, bottom panel. Since Ba-abundances are very sensitive to the stellar parameters, in particular to the microturbulence velocity, we tested whether we would get similar s-process enhancements for our stars if we exclude the Ba-abundance. We performed again the same $\chi^2$ tests to the abundances of all Ba-normal stars, averaged together and the two Ba-rich stars - separately. The results show that the Ba-normal stars have again experienced only $10\%$ of the solar s-process and the two Ba-rich ones $40\%$ and $70\%$, respectively.

Since the s-process enhancement is usually associated with AGB stars, this can be seen as evidence that FRMS were the main polluters, which ejecta formed the intermediate stellar population. One should note, however, that only the most massive ($\sim5 - 8 M_{\odot}$) AGB stars reach the necessary high ($>8\times10^7$~K) temperatures at the bottom of the convective envelope to activate the ON cycle and thus to reduce the O abundance. These AGB stars experience only a few dredge-up processes, which cannot alter the s-process abundances \citep{dercole+2008}. Thus, the AGB scenario cannot be ruled out. In fact, it is even more favourable, considering the presence of 2 Ba-rich stars, discussed in the previous section. In conclusion, the heavy elements were not significantly modified during the self-enrichment processes in the early cluster evolution (with the noted exception of the few most Na-rich stars) but their pattern is genuine to the cloud, from which M~75 formed. 



\subsection{Th-Eu age}

Despite the large formal errors of the measured Th abundance, the star-to-star variations of Th are small, which leads to a precise, mean Th abundance for the cluster. We used the [Th/Eu] ratio to derive an approximate age estimate for M~75 based on the radioactive decay of Th. The mean $\log\epsilon(\rm{Th/Eu})$ is $-0.55\pm0.02$~dex, or $N_{\rm Th}/N_{\rm Eu}=0.28$. Whilst Eu is a stable element, the half life period ($t_{1/2}$) of the isotope $^{232}\rm{Th}$ is $1.41\times10^{10}$~yr. Using the universal law of radioactive decay, namely $N(t)=N_0 e^{-\lambda t}$, where $\lambda = \ln(2)/t_{1/2}$, and the Solar System initial Th/Eu ratio, $N_{\rm Th}/N_{\rm Eu}=0.46$, \citep{cowan+1999}, we obtained an age estimate of $10$~Gyr. This method usually leads to a precision of $\pm3$~Gyr but the estimate  we reached for M~75 is in perfect agreement with isochrone ages for this object \citep[e.g.][]{catelan+2002} and confirms that it is younger than the bulk of MW GCs.
For comparison, \citet{sneden+2000} have published $N_{\rm Th}/N_{\rm Eu}=0.25$ for M~15, which leads to an age of $12.4$~Gyr, if using the same atomic parameters. Note that very small systematic errors in the Th abundance can still lead to large errors in the age estimate. For instance, a systematic error of $0.1$~dex in Th leads to $5$~Gyr difference in age. More accurate spectroscopic age-dating would be feasible once U-abundances can be measured \citep{frebel+2007}, but given the low S/N ratios in the relevant blue regions, this is an unlikely endeavour in the remote M~75.



\subsection{Multiple populations in the CMD of M~75}

The phenomenon of multiple populations in GCs often extends to the appearance of the CMD. High-precision photometry has revealed multiple main sequences, subgiant branches, and RGBs in the CMDs of many clusters, which do not exhibit large variations in metallicity \citep{piotto+2007,piotto+2012,han+2009}. These effects are mostly driven by CNO variations and different He-content in stars from different generations \citep{dantona+2002,dantona+2005,piotto+2005}. Moreover, the presence of multiple populations has been proven to be one of the key parameters that shapes the HB, where the effect is most pronounced, because stars of the same age but different He-content have different initial masses and, thus, occupy different regions of the HB.
Whilst the stars with primordial He-abundance (compatible with the Big Bang nucleosynthesis) preferably populate the red part of the HB, stars with enhanced He are responsible for the formation of extended blue tails \citep{dantona+2002,dantona+caloi2004}.
Strong correlations between the He-content and the abundances of p-capture elements with the effective temperature amongst stars from the HB have been recently found in M~4 \citep{marino+2011}, NGC~2808 \citep{gratton+2011}, and NGC~1851 \citep{gratton+2012b}.
M~75 is one of the most curious cases in this respect. It has a very extended and peculiar HB with a trimodal distribution \citep{catelan+2002}. Apart from the well separated BHB and RHB, its CMD shows a distinct third extension of a very blue, faint tail \citep[Figure \ref{fig:CMD}, see also Figure 2 in][]{catelan+2002}. The extended blue tail of M~75's HB is at odds with our findings of only a moderate Na-O anticorrelation on the RGB (Figure \ref{fig:anticor}), which so far lacks an extreme (E) population.
Such a population is often found in GCs with extended HBs \citep{carretta+2009b} and characterised by extremely Na-rich and O-poor stars, and accompanied by large He variations. Having in mind, however, the tiny populated extremely blue tail in M~75 and the limited number of our sample, it is possible that such E population has just been missed by our selection criteria. But curiously, we do not detect a Mg-Al anticorrelation, neither, which is commonly found in those GCs with a more complex HB morphology. Furthermore, clues for He-, CNO-, or age-variations have not been detected in the CMD of M~75 (in terms of multiple RGBs, subgiant branches, or main sequences). We note however, that there is not any narrow-band photometry available for this GC, which might reveal multiple populations amongst the main sequence, the subgiant branch and the RGB \citep[e.g.][]{carretta+2011b}.
Still, the distribution of He and p-capture elements amongst the HB of M~75 remains an open question and thus, high-resolution spectroscopic observations amongst stars on the HBs are needed to ascertain if its peculiar morphology is mostly driven by the presence of multiple populations or if there are other parameters with major influence.
Possibly, the younger age and higher concentration, complemented with its remote location in the Milky Way halo hold important clues about its origin.

\subsection{Four distinct chemical populations in M~75}

A closer look to the Na-O and Al-O anticorrelations of M~75 (Figure \ref{fig:anticor}) shows three distinct populations rather than a continuous anticorrelation, as found in most GCs (the P-population, which consists of stars with Na- and O-abundances typical of the halo field; a second group of stars mildly enriched in Na and depleted in O; and a third group of the most Na-rich, O-depleted stars, which is separated by a clear gap from the second group).
But the question whether the Na-O anticorrelation in GCs is actually continuous or rather discrete has recently raised attention because the findings of discrete main sequences in some GCs \citep{dantona+2005,piotto+2007} and the discrete distribution of the Na and O abundances amongst the HB \citep{marino+2011} suggest a discrete chemical distribution of the different populations also amongst the RGB. Currently, more precise, high-resolution studies of large number of GCs' RGB stars are being carried out to clear out this question \citep[e.g.][]{carretta+2012}. Besides the three populations seen in the Na-O plane, M~75 also hosts a number of Ba-rich stars, which could represent a fourth population.

\begin{figure}
\centering
\resizebox{\hsize}{!}{
\includegraphics[angle=0]{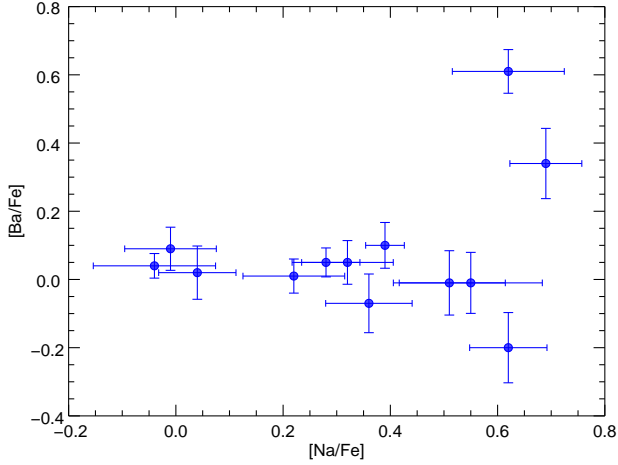}
}
\caption[]{[Ba/Fe] vs. [Na/Fe] ratios in M~75's RGB stars. The two most Na-rich stars are also Ba-enhanced by $0.34\pm0.10$~dex and $0.61\pm0.06$~dex, respectively.
}
\label{fig:BaFe}
\end{figure}

A possible explanation of the formation of four populations could be found within the pristine gas dilution scenario suggested by \citet{dercole+2008,dercole+2011}. On a timescale of several tens of Myr after the P-generation has formed, the centre of the cluster still hosts only gas from the higher mass AGB stars' ejecta. At this point, the Na-rich, O-depleted, Ba-normal population is formed.  In the next $\sim10$~Myr, pristine gas falls in the central region of the cluster and mixes with the gas enriched by the AGB winds. Another population forms from the diluted gas, which is mildly Na-rich and O-depleted. After the diluted gas is fully processed, the lower mass AGB stars remain the only source of gas in the cluster. Their ejecta could also be enhanced in s-process elements. The last Ba-rich population forms, which is also strongly Na-rich and O-depleted and indistinct from the other stars formed from none-diluted gas (Figure \ref{fig:BaFe}).
Hydrodynamical and N-body simulations can test the viability of this idea and better constraint the time-scales and polluters' masses. Meanwhile, some improvements of the AGB models are also needed, since we do not yet fully understand these very late evolutionary stages in terms of mass-loss, convection, and nuclear reaction cross-sections.

\subsection{Constraints on the mass of the polluters in M~75}

\begin{figure}
\centering
\resizebox{\hsize}{!}{
\includegraphics[angle=0]{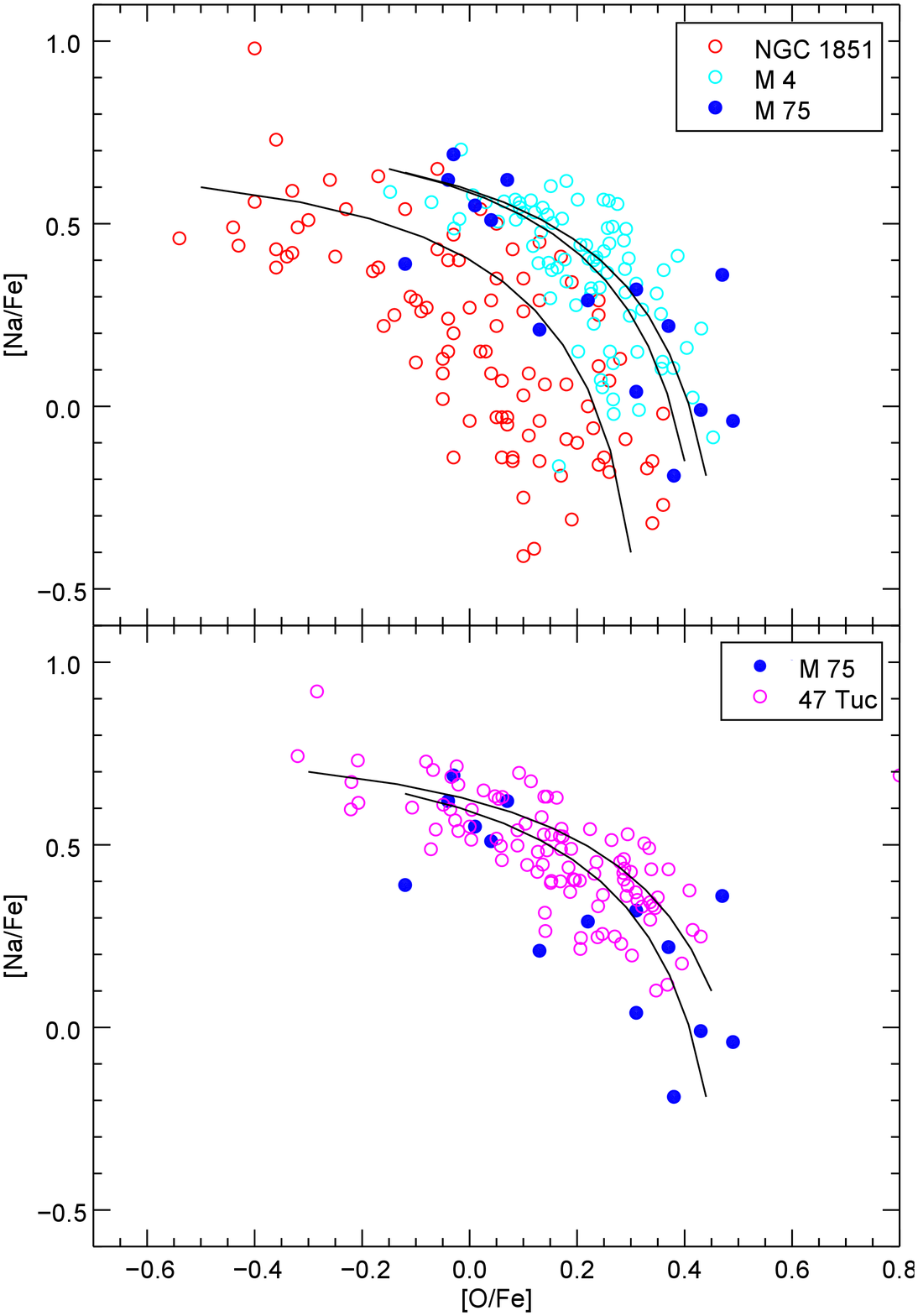}
}
\caption[]{A comparison of the extent of the Na-O anticorrelations of several GCs relative to M~75; Top panel: the Na-O anticorrelations in M~75, NGC~1851 \citep{carretta+2011}, and M~4 \citep{carretta+2009b}; Bottom panel: the Na-O anticorrelations in M~75 and 47~Tuc \citep{carretta+2009b}. Simple dilution models are overimposed. The [O/Fe] abundances of M~75 are shifted by the mean discrepancy of [Fe I/Fe II]$=-0.14$~dex to match the other studies.
}
\label{fig:NaO}
\end{figure}

Now we can place some qualitative constraints on the average mass of the polluters in M~75 provided that enrichment was dominantly from AGB stars. Despite its high luminosity, which is often considered as a reason for the presence of higher mass polluters, there are several factors prompting that {\em lower} mass AGB stars also contributed to the formation of the intermediate population in this GC: 

i) Whilst the maximum amount of Na produced in most GCs is approximately the same, high mass AGB stars manage to process more O through the ON cycle and, hence, they are more depleted in this element. 
Therefore, the amount of O-depletion is a proxy for the polluters' average mass in the sense that more extended Na-O anticorrelations imply higher mass polluters \citep[see][]{carretta+2009b}.
For instance, the Na-O anticorrelation in M~75 is less extended than in other GCs of similar luminosity (e.g. NGC~1851, NGC~2808, M~5) and more similar to the less massive M~4, prompting for lower mass polluters in M~75.
Another example of a massive GC that does not have an extended Na-O anticorrelation is 47~Tuc (M$_{\rm{V}}=-9.42$~mag). This GC, however hosts significantly more O-depleted stars compared to M~75 and suggests predominantly high-mass AGB polluters, in accordance with the conclusions drawn by \citet{carretta+2012}. This is best represented in Figure \ref{fig:NaO}, where we plot the Na-O anticorrelations of M~75 and the three reference clusters -- NGC~1851, M~4, and 47~Tuc. We also show simple dilution models that we computed as described in \citet{carretta+2009b}, tuning by eye the input parameters (the maximum and minimum Na- and O-abundance ratios) for all of them. 
Figure \ref{fig:NaO} clearly shows that the Na-O anticorrelation of M~75 is more similar to the less massive GC M~4 compared to the extended anticorrelation of NGC~1851 and ``steeper'' than the Na-O anticorrelation of 47~Tuc. Thus, we suggest that the average mass of the polluters differs from cluster to cluster, regardless of its luminosity. M~75 and M~4 were likely enriched by the ejecta of lower mass AGB stars, whilst 47~Tuc was mainly enriched by more massive AGB stars. GCs with very extended Na-O anticorrelations (e.g. M~5, M22, NGC~2808, NGC~1851) most likely experienced continuous star formation ($\sim100$~Myr) and were enriched by polluters of broader mass range.

ii) Furthermore, the two stars, which present anomalously high Ba abundances are also the most Na-rich ones (Figure \ref{fig:BaFe}). This is indicative of s-process enrichment from intermediate mass AGB stars ($\rm{M} < 4 - 5 \rm{M}_{\odot}$), that are able to alter the s-process pattern of the I-population stars \citep{gallino+1998}. Ba-rich stars are very rare in GCs but have also been found in NGC~1851, M~22, and $\omega$~Cen, which are all noted for their complex evolution, likely characterised by longer star formation period extending after these low mass stars manage to pollute the cluster environment with s-process enhanced ejecta.

iii) The [Rb/Zr] ratio can also be a proxy for the polluters' mass because it is sensitive to the neutron density. The main neutron source in the He shell of low mass ($1 <\rm{M} < 4 \rm{M}_{\odot}$) AGB stars is the reaction $^{13}\rm{C}(\alpha,n)^{16}\rm{O}$, whilst in more massive AGB stars, neutrons are mainly released by $^{22}\rm{Ne}(\alpha,n)^{25}\rm{Mg}$ and since the $^{22}$Ne source produces much higher neutron densities than the $^{13}$C neutron source, the [Rb/Zr] ratio can discriminate between the two \citep{garcia-hernandez+2009}. The mean [Rb/Zr] ratio of our sample of M~75 stars is $=-0.04\pm0.03$~dex and there are not any significant star-to-star variations in our sample of stars. Both Ba-rich stars have a $\rm{[Rb/Zr]}=0.06\pm0.2$~dex, which is a statistically insignificant difference with respect to the mean [Rb/Zr] ratio. We conclude that we cannot use the [Rb/Zr] ratio to discriminate between different AGB masses, owing to the large uncertainties of this ratio in individual stars.



\subsection{Comparison with other GCs, Galactic halo field stars, and dwarf spheroidal galaxies}

\begin{figure}
\centering
\resizebox{\hsize}{!}{
\includegraphics[angle=0]{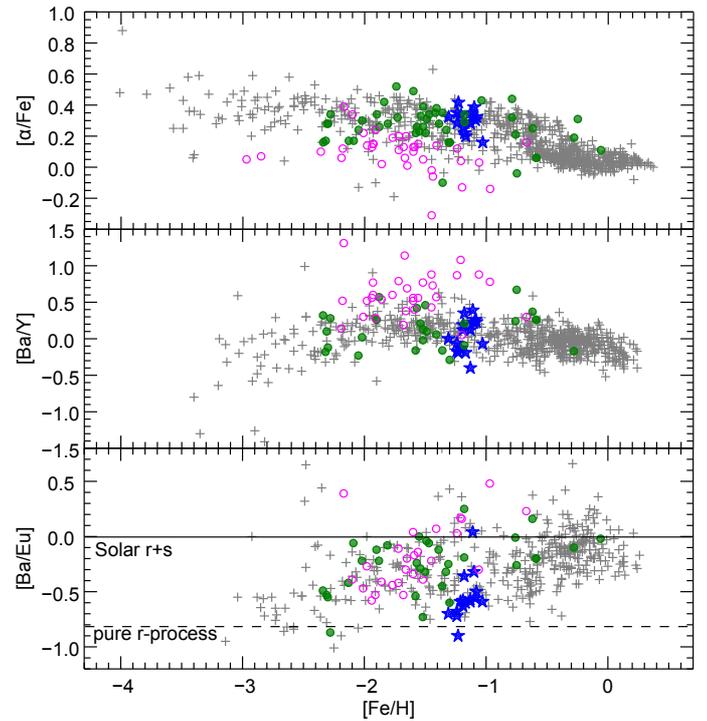}
}
\caption[]{A comparison of the $\alpha$ and n-capture element abundances of the 16 stars of M~75 (blue asterisks) with Galactic disk and halo stars (grey crosses), average abundance values of other Galacatic GCs (filled green circles), and a representative sample of individual dSph stars (open magenta circles).
}
\label{fig:abundance_comp}
\end{figure}

In order to investigate M~75's origin, we present in Figure \ref{fig:abundance_comp} a comparison of the abundances in M~75 with the abundances of Galactic disk and halo stars at different metallicities, and with the average abundances of other GCs. A small representative sample of individual dSph stars is also plotted for comparison.
The abundances of the Milky Way halo and disk stars are taken from the compilation of \citet[][and references therein]{venn+2004} and complemented with the recent results of \citet{ishigaki+2012a,ishigaki+2012b}. 
The sample of dSph stars also comes from \citet{venn+2004} and includes the abundances of individual stars from Carina, Fornax, Leo, Scl, UMi, Sex, and Draco dSphs.
The mean abundances of various GCs are taken from \citet{pritzl+2005} and complemented with the more recent results for NGC~1851 \citep{carretta+2011}, M~5 \citep{yong+2008}, and the outermost halo clusters Pal~3 \citep{koch+2009}, and Pal~4 \citep{koch+cote2010}.

We chose to plot in Figure \ref{fig:abundance_comp} three key element ratios, important to trace the chemical evolution of M~75. The $\alpha$-element abundance ratio in M~75 is fully compatible with the Galactic halo stars at the same metallicity and consistent with the $\alpha$-enhanced old stellar populations of the Milky Way halo. This suggests that it experienced the same (fast) star formation history, dominated by SNe II and only late, delayed Fe enrichment by SNe Ia, as most MW GCs. A connection with the dSph galaxies and their low star formation rates (hence low [$\alpha$/Fe]) can be ruled out.
The latter scenario has been suggested for some GCs with low $\alpha$-abundance like Pal~12, Ruprecht~106, and possibly Ter~7, associated with the Sgr dwarf \citep{pritzl+2005}.

The [Ba/Y] ratio compares the abundance yields between the main s-process, which takes place in intermediate- to low-mass AGB stars and the weak s-process, which is associated with very massive ($M>20~M_{\odot}$) stars \citep{burris+2000}. The Galactic halo and disk stars have a roughly solar [Ba/Y] ratio over a broad range of metallicities, which starts to decrease at [Fe/H]$~\lesssim-2$~dex. In dSphs, on the other hand, the [Ba/Y] ratio is rather high \citep{venn+2004,tolstoy+09}, owing to their typically lower star formation rates. However, all GCs, including M~75, present typical of the halo field [Ba/Y] ratios around the solar value, with again, the notable exception of Pal~12, which has unusually high mean [Ba/Y] ratio, more similar to the stars from dSph galaxies.

Finally, the [Ba/Eu] ratio compares the yields from the main s- and main r- neutron capture processes. The latter operates mainly in massive stars during the eruptions of SNe II. Typically, all stars in the field and in GCs present abundances that are consistent with r-process enrichment plus some fraction of the solar s-process contribution from AGB stars. The s-process fraction varies for different stars but the general trend is that it could be entirely missing for metal poor stars and rises until it becomes dominant for the metal rich population. As we noted above, M~75 has an unusually low, mean s-process contribution for its metallicity, but its [Ba/Eu]$=-0.55\pm0.05$~dex ($-0.63\pm0.01$~dex if we consider only the P-generation) is still consistent with some halo stars. Only a few GCs, studied to date, have been noted to present [Ba/Eu] ratios compatible or lower than M~75.
The majority of them are amongst the most metal poor GCs in our Galaxy with [Fe/H] below $-2.3$~dex, namely M~68, [Ba/Eu]$=-0.50$ \citep{lee+2005}, M~92, [Ba/Eu]$=-0.55$ \citep[][]{shetrone+2001}, M~15, [Ba/Eu]$=-0.87$ \citep[][]{sneden+2000}, and M~30, [Ba/Eu]$=-0.53$ \citep[][]{shetrone+2003}. On the more metal-rich end, we note the GCs NGC~3201 with [Fe/H]$=-1.58$~dex and [Ba/Eu]$=-0.54$~dex \citep{gonzalez+wallerstein1998}, Pal~3 with [Fe/H]=$-1.52$~dex and [Ba/Eu]$=-0.73$~dex \citep{koch+2009}, and M~5 with [Fe/H]$=-1.30$~dex (most similar to M~75) and [Ba/Eu]$=-0.60$~dex \citep{ramirez+cohen2003,yong+2008}. 

NGC~1851 is probably the GC that shares most common properties with M~75 and is often thought as its twin \citep{catelan+2002}. Here, we investigate the similarities and differences between the two objects in deeper detail. Both GCs are coeval, share the same metallicity, and show similar HB morphology. Both are luminous, massive and very concentrated clusters located in the transition region between the inner and outer Milky Way halo. A notable difference between the two objects is the presence of double subgiant and red giant branches in NGC~1851 \citep[see][]{milone+2008, han+2009}, which are not observed in M~75, despite the same photometric quality. \citet{ventura+2009} have suggested that large CNO variations and a small age spread could explain the subgiant branch in NGC~1851. This scenario is supported by \citet{yong+2009} from observations of 4 RGB stars.
However, it has been dismissed by more recent studies by \citet{villanova+2010} of 15 RGB stars and by \citet{gratton+2012b}, who derived abundances for a large sample of HB stars. Both studies found no evidence for significant CNO variations. \citet{gratton+2012b} concluded that the only explanation of the splitting of the SGB and RGB of NGC~1851 is a considerable age difference of about $1.5$~Gyr. Furthermore, \citet{carretta+2010b} detected two independent Na-O anticorrelations on the two RGBs of NGC~1851, which led to the suggestion that this cluster could have formed from the merger of two GCs with different ages. In any case, the clearly more complex formation history of NGC~1851 is responsible for the very extended HB in this system. In the case of M~75, it is not yet clear what physical processes drive the formation of such an extended HB.

\begin{table*}
\begin{center}
\caption{Comparison with GCs that share some common properties with M~75}\label{tab:GCs}
{\normalsize
 \begin{tabular}{ccccccc}
\hline
     & M~75 & NGC~1851 & M~4 & M~5 & 47~Tuc & Pal~3 \\
\hline
R$_{\rm{GC}}$ (kpc)\tablefootmark{1} & 14.7 & 16.6 & 5.9 & 6.2 & 7.4 & 95.7 \\
r$_{\rm{h}}$ (arcmin)\tablefootmark{1} & 0.46 & 0.51 & 4.33 & 1.77 & 3.17 & 0.65 \\
$c = \log(r_t/r_c)$\tablefootmark{1}  & 1.80 & 1.86 & 1.65 & 1.73 & 2.07 & 0.99 \\
rel. age (Gyr)\tablefootmark{2} & $\sim10$ & $9.8$ & $12.7$ & $10.8$ & $13.7$ & ? \\
M$_{\rm{V}}$\tablefootmark{1} & $-8.57$ & $-8.33$ & $-7.19$ & $-8.81$ & $-9.42$ & $-5.6$ \\
$\rm{[Fe/H]}$\tablefootmark{1} & $-1.16$ & $-1.18$ & $-1.16$ & $-1.29$ & $-0.72$ & $-1.63$ \\
$\rm{[\alpha/Fe]}$\tablefootmark{3} & 0.30 & 0.34 & 0.29 & 0.16 & 0.25 & 0.39 \\
$\rm{[Ba/Eu]}$\tablefootmark{3} & $-0.63$ & $-0.19$ & $0.25$ & $-0.60$ & $0.16$ & $-0.73$ \\
$\rm{[Ba/Y]}$\tablefootmark{3} & $0.04$ & $0.21$ & $-0.09$ & $-0.29$ & $0.37$ & $-0.02$ \\
\hline
 \end{tabular}
\par}
\tablefoot{
\tablefoottext{1}{Data taken from \citet[][2010 version]{harris96}}
\tablefoottext{2}{Relative ages from \citet{marin-franch+2009}. We adopted a reference age of 13~Gyr.}
\tablefoottext{3}{Various sources cited throughout the text.}
}
\end{center}
\end{table*}

\begin{figure}
\centering
\resizebox{\hsize}{!}{
\includegraphics[angle=0]{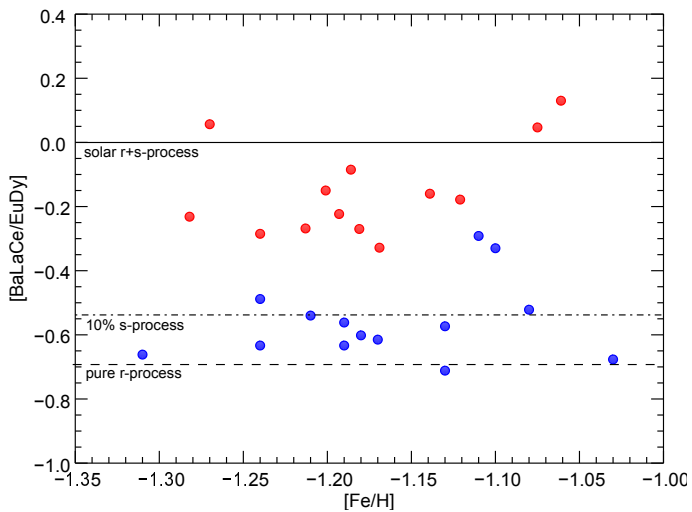}
}
\caption[]{A comparison between the r- and s-process enrichment in M~75 (blue dots) and NGC~1851 (red dots). The figure clearly reveals the different enrichment history between the two GCs.
}
\label{fig:m75_n1851_rs}
\end{figure}

The most comprehensive chemical study of NGC~1851, in terms of derived abundances for many different elements, is presented in \citet{carretta+2011}. We used it to compare the chemical abundances in both clusters. They share the same metallicity (with no evidence of significant iron spreads in neither of them) and similar $\alpha$-enhancement. The p-capture elements Na and Al have similar variations in both clusters but O shows a larger spread in NGC~1851, leading to a more extended Na-O anticorrelation in the latter GC (See Figure \ref{fig:NaO}). 
The iron-peak element abundances are identical in both GCs. The largest difference between the two lies in the n-capture elements. This is best illustrated in Figure \ref{fig:m75_n1851_rs}, where we present the total s- to r-process ratio in both GCs. We chose the average abundance of Ba, La, and Ce as representatives of typical s-process elements, where about $80\%$ of their production comes from the s-process, and the average abundance of Eu and Dy as representatives of elements produced mainly by the r-process \citep{burris+2000}.
The stars of NGC~1851 clearly lie above those from M~75 in this parameter space, indicating different primordial s-process contribution. We note, however, that both M~75 and NGC~1851 host s-process rich stars, which were enhanced in s-process elements most probably by intermediate mass AGB stars during the early evolution of these stellar systems via self enrichment mechanisms, but the s-process rich stars in M~75 reach [s/r] ratios similar to the primordial s-process enhancement of NGC~1851.

Finally, we note that M~75 is both a unique and a normal GC of the Milky Way's GCs system. Unique, in the sense that there is not a GC, which resembles all the same properties of M~75, and normal in the sense that its properties fit well in the general picture of the Milky Way's GCs. So far there are not two clusters found to be exactly alike and each one of them deserves special attention. In Table \ref{tab:GCs} we present some important characteristics of M~75 compared to other GCs, which were discussed in this section and that share some important similarities and differences with M~75.

\section{Summary}

In this work, we presented the first chemical abundance study of the outer halo GC M~75. Our data sample consists of high resolution spectra of 16 giant stars, obtained with the MIKE spectrograph at the Magellan Observatory. We derived abundances through EW measurements and spectral synthesis in LTE for a total of 32 different elements covering a broad range of p-capture, $\alpha$, iron-peak, and n-capture elements. M~75 is moderately metal rich cluster with [Fe/H]$=-1.16\pm0.02$~dex with a marginal spread of $0.07$~dex, typical for GCs with similar luminosity.
We measured an enhanced average $\alpha$ abundance [$\alpha$/Fe]$=0.30\pm0.02$~dex, based on Mg, Si, and Ca, typical for the Galactic halo at this metallicity. We found significant variations in the abundances of the p-capture elements O, Na, and Al, which provide evidence for the presence of at least two generations, formed on a short time-scale. Sodium is anticorrelated with O and correlated with Al, consistent with simple dilution models.
The Na-O anticorrelation appears discrete, suggesting three chemically distinct populations. Additionally, the two most Na-rich stars form a fourth, Ba-enhanced population. Based on the extent of the Na-O anticorrelation, we conclude that the I-population stars were enriched by the ejecta of relatively less massive AGB stars in several episodes of star formation, which ended before the SNe Ia began to contribute iron to the cluster's environment. We note that the least massive polluters were able to alter the s-process abundances of the cluster's ISM. 

The moderate O-Na anticorrelation (our sample of 16 stars lacks an extreme population of stars with very low O abundances) and the lack of significant Mg variation are at odd with the very extended trimodal HB of M~75. We conclude that the parameters that shape the peculiar HB morphology of this GC are still unclear and more observations are required, in particular a spectroscopic sample of stars, which represents the full span of the HB. A careful CNO abundance analysis of the existing spectroscopic sample is foreseen in a subsequent paper (Kacharov et al., in prep.).

The n-capture element pattern is consistent with predominant r-process enrichment with a marginal contribution (about $10\%$ of the scaled solar production) of s-process. 

The overall chemical, evolutionary status of M~75 is consistent with other inner and outer halo GCs and field stars, which suggest a similar origin with the bulk of Milky Way GCs. Despite its large galactocentric distance, coupled with its high metallicity and younger age, M~75 does not seem to present any odd, chemical properties, that would indicate extragalactic origin and accretion to the Milky Way halo at a later stage of its evolution.

\begin{acknowledgements}
The authors thank Miho Ishigaki for providing tables with abundances of heavy elements in the Galactic halo and Raffaele Gratton and Francesca D'Antona for helpful discussions on multiple populations in GCs.
NK and AK acknowledge the Deutsche Forschungsgemeinschaft for funding from  Emmy-Noether grant  Ko 4161/1.
\end{acknowledgements}

\bibliographystyle{aa}

\bibliography{mybiblio_v3}


\end{document}